\documentstyle[epsfig,amsmath,times]{mn}
\def\spose#1{\hbox to 0pt{#1\hss}}
\def\oo{[O\textsc{ii}]}
\def\ooo{[O\textsc{iii}]}
\def\lta{\mathrel{\spose{\lower 3pt\hbox{$\mathchar"218$}}\raise 2.0pt\hbox{$\mathchar"13C$}}}
\def\gta{\mathrel{\spose{\lower 3pt\hbox{$\mathchar"218$}}\raise 2.0pt\hbox{$\mathchar"13E$}}}
\def\arcsec{$^{\prime\prime}$}
\def\crat{C\textsc{iii}]1909/C\textsc{ii}]2326 }
\def\nrat{[Ne\textsc{iii}]3869/[Ne\textsc{v}]3426 }

\title[Spectroscopy of $\mathbf{z \sim 1}$ 6C radio galaxies - I]
{Deep spectroscopy of $\mathbf{z \sim 1}$ 6C radio galaxies - I. The effects of
radio power and size on the properties of the emission line gas.}
      
\author[K.\,J.\, Inskip {\it et al.}]
{K.\,J.\, Inskip$^1$\footnotemark, P.\,N.\,Best$^2$, S.\,Rawlings$^3$,
M.\,S.\,Longair$^1$, G.\,Cotter$^1$,\cr H.\,J.\,A.\,R\"{o}ttgering$^4$
and S.\,Eales$^5$\\  
$^1$ Cavendish Laboratory, Madingley Road, Cambridge, CB3 0HE,\\ $^2$
Institute for Astronomy, Royal Observatory Edinburgh, Blackford Hill,
Edinburgh, EH9 3HJ\\$^3$ Department of Astrophysics, University of
Oxford, Keble Road, Oxford OX1 3RH\\ $^4$ Sterrewacht Leiden, Postbus
9513, 2300 RA Leiden, the Netherlands\\ $^5$ Department of Physics and
Astrophysics, University of Wales Cardiff, PO Box 913, Cardiff CF2 3YB\\ }

\date{ }

\pagerange{\pageref{firstpage}--\pageref{lastpage}}

\pubyear{2002}

\begin{document}

\label{firstpage}
\maketitle
\begin{abstract}
The results of deep long-slit optical spectroscopy for a sample of eight
6C radio galaxies at redshift $z\sim 1$ are presented. Emission line
ratios are derived for many emission lines with rest--frame wavelengths of
$1500 - 4500$\AA\, and the kinematic properties of the emission line gas
are derived from an analysis of the two dimensional structure of the 
[O\textsc{ii}]3727\AA\, emission line at $\approx 5\rm{\AA}$ spectral 
resolution.  

In general, the 6C spectra display many characteristics similar to those
of more powerful 3CR sources at the same redshifts. The emission line
region gas kinematics are more extreme for the smaller radio sources in
the sample, which often display distorted velocity profiles. The 
ionization state of the emission line region also varies 
with radio size: the spectra of large radio sources ($> 120$kpc) are 
consistent with photoionization by an obscured AGN, whilst smaller ($< 
120$kpc) sources typically exist in a lower ionization state and have
spectra which are better explained by additional ionization due to shocks
associated with the expanding radio source. The kinematic and ionization
properties of the 6C radio galaxies are clearly linked. As for the 3CR 
sources, smaller radio sources also typically possess more extensive 
emission line regions, with enhanced emission line luminosities. A high 
velocity emission line gas component is observed in 6C1019+39, similar to 
that seen in 3C265.   

It is clear that the best interpretation of the spectra of radio
sources requires a combination of ionization mechanisms. A simple
model is developed, combining AGN photoionization with photoionization
from the luminous shock associated with the expanding radio
source. The relative contributions of ionizing photons from shocks and
the central AGN to an emission line gas cloud varies with radio source
size and the position of the cloud. This model provides a good
explanation for both the ionization properties of the emission line
regions and the radio size evolution of the emission line region
extents and luminosities.  

\end{abstract}


\section{Introduction}

\footnotetext{E-mail: kji@mrao.cam.ac.uk}

Extended emission line regions are often observed around powerful radio galaxies.
High redshift ($z \gta 0.5$) radio galaxies typically have more
extensive emission line regions than lower power radio sources at low
redshift. These emission line regions can be up to 100\,kpc in size,
and are often elongated and aligned along the radio jet axis (McCarthy
1993;  McCarthy {\it et al} 1995).  The kinematics of the emission
line gas of higher redshift radio galaxies are generally more extreme
than those seen in low redshift radio galaxies; more distant sources
have larger emission line FWHM, and larger velocity shears. The
luminosities of the emission lines are large, with rest frame
equivalent line widths often exceeding 100\AA\, (e.g. Baum \& McCarthy
2000).   

Over the past few years it has become increasingly clear that in
addition to photoionization by an obscured AGN, in some cases the
emission line gas may also be ionized by radiative shocks associated with the
radio source.  Evidence for this includes the fact that whilst the
spectra of low redshift sources are well explained by AGN
photoionization, the emission line regions of more distant galaxies do
not always display the characteristic line ratios of photoionized
regions, and require alternative heating mechanisms.  Furthermore,
observations of individual sources reveal features suggesting
significant interactions of the radio jet with the interstellar medium
(ISM) of the host galaxy.  Shocks associated with the passage of the
radio source clearly influence the kinematics and morphology of the
emission line gas, and can also affect the ionization state within the
gas.    
Another potential source of ionizing photons is UV emission from a
young stellar population.   Dey {\it et al} (1997) find that the
extended UV continuum emission from 4C 41.17 (z = 3.8), is
unpolarised, suggesting that scattered light from the AGN is not a
dominant source of UV photons.  This galaxy appears to have undergone
a major epoch of star formation at $z \sim 4$.    At lower redshifts,
the stellar 
populations of radio galaxies are generally old.  Most 3CR radio
galaxies at $z \sim 1$ exhibit strong continuum polarisation ($\sim
10\%$, e.g. di Serego Alighieri, Cimatti \& Fosbury 1994); scattered
light from an obscured AGN provides a large proportion of the extended
UV continuum.  Although small amounts of radio source jet induced star
formation are possible, this is clearly not the dominant source of
aligned continuum emission, nor can it be a major source of ionizing
photons for objects at $z \sim 1$. 

Spectroscopic observations of a sample of 14 3CR radio galaxies at $z
\sim 1$ by Best {\it et al.} (2000a, 2000b) showed that the ionization
state of such sources is strongly correlated with radio size, smaller
radio sources generally existing in a lower ionization state. In
addition, radio sources with linear sizes $\lta 150$kpc typically have
greater emission line fluxes and broader line widths than their larger
counterparts. The emission line ratios of smaller sources are in good
agreement with the predictions of shock ionization models, and their
observed kinematics are more disturbed. The irregularities in the
velocity structures of the small sources are likely to be due to shock
acceleration of the emission line gas clouds. Compression of the
emission line gas clouds by the shock front and the  ionizing photons
associated with it combine to lower the ionization state of the gas in
smaller sources.  The properties of the larger, older, radio sources,
for which the shocks associated with the expansion of the radio cocoon
are long in the past, are well explained by photoionization models. 

A spectroscopic study of four intermediate redshift 3CR galaxies
by Sol\'{o}rzano-I\~{n}arrea {\it et al.} (2001) found evidence for
shock acceleration of the emission line gas in the extended emission
line regions (EELRs) of all four galaxies. The disturbed kinematics of
these sources are reflected in the emission line spectra by line
splitting and/or underlying broad components.  Two of these sources 
have radio sizes comparable to the physical extent of the emission
line region, and both show evidence for an underlying broad component
in their emission lines.  Of the two sources with radio sizes
significantly larger than their emission line regions, only one shows
evidence for an underlying broad component, although both sources
exhibit line splitting.  Both of these features can be related to 
the effects of shocks on the emission line gas. 

It is clear that the age of a radio source and the dominant ionization
mechanism play a determining role in the observed properties of the
emission line regions of powerful high redshift radio galaxies.  It is
important to understand exactly how radio luminosity correlates with
the EELR properties; the precise effects of radio jet power on shock
acceleration and the ionization mechanism are still uncertain.  To this
end, we have carried out deep spectroscopic observations of a complete
sample of 8 6C radio galaxies at $z \sim 1$.  In this paper we
investigate the effects of radio power on the ionization and kinematic 
properties of emission line regions by contrasting the properties of
our sample of 6C galaxies with those of the more powerful 3CR galaxies
at similar redshifts.   A comparison with a sample of low redshift 3CR
sources matched in radio power to the 6C subsample will also enable us
to  break the degeneracy between redshift and radio power present in
any flux limited sample; this analysis is deferred to a second paper.
Overall, this research is aimed at adding to our understanding of: (i)
the physics of the ISM in relatively extreme conditions, (ii) the
impact of AGN activity on its nearby environment and (iii) the origin
of the cool gas which forms the extended emission line regions. 

\begin{table*}
\caption{Details of the ISIS observations. Where more than one slit
position angle
is listed on a single line, the radio galaxy emission line data were
combined in 2--dimensions, due to the similarity of the 2--d emission
line spectra.}  
\begin{center} 
\begin{tabular} {lccccccc} 
\hline
{Source}&{Redshift}&{Observation}&{Slit width}&{Isis Arm}&{Exposure}&{Wavelength}&{Slit PA}\\ 
{}&{}&{Date}&{[arcsec]}&{}&{time [s]}&{Range [\AA]}&{[deg.]}\vspace*{0.05cm}\\\hline 
6C0943+39 & 1.035 & 21/03/99, 22/03/99           & 1.5     & blue & 15360$^1$  & 3100--5400 & 105$^2$\\
          &       & 21/03/99, 22/03/99, 01/03/00 & 1.5-2.5 & red  & 19950  & 6900--9000 & 105, 110$^{2}$\vspace*{0.15cm}\\
6C1011+36 & 1.042 & 21/03/99           		 & 1.5     & blue & 3000   & 3100--6000$^3$ & 47$^2$\\
          &       & 21/03/99           		 & 1.5     & red  & 3000   & 6900--8250 & 47$^2$\\
	  &   	  & 21/03/99, 01/03/00 		 & 1.5-2.5 & blue & 9320   & 3100--6000 & 345$^4$\\
          &   	  & 21/03/99$^5$   	         & 1.5-2.5 & red  & 9000   & 6900--9000 & 345$^4$\vspace*{0.15cm}\\
6C1017+37 & 1.053 & 22/03/99, 29/02/00 		 & 1.5-2.5 & blue & 15440  & 3100--5400 & 48$^2$\\
          &    	  & 22/03/99, 29/02/00 		 & 1.5-2.5 & red  & 13180  & 6900--9000 & 48$^2$\vspace*{0.15cm}\\ 
6C1019+39 & 0.922 & 28/02/00           		 & 1.5-2.0 & blue & 9360$^1$   & 3100--5400 & 61$^2$\\
          &       & 28/02/00, 01/03/00 		 & 1.5-2.5 & red  & 14800  & 6400--7900$^6$ & 61$^2$\vspace*{0.15cm}\\
6C1129+37 & 1.060 & 21/03/99, 22/03/99 		 & 1.5     & blue & 12590  & 3100--5400 & 105$^2$\\
          &       & 21/03/99, 22/03/99 		 & 1.5     & red  & 12000  & 6950--8450$^7$ & 105$^2$\vspace*{0.15cm}\\
6C1217+36 & 1.088 & 29/02/00           		 & 2.0     & blue & 10540  & 3500--5400 & 61$^2$\\
          &       & 29/02/00           		 & 2.0     & red  & 7300$^8$   & 7100--8600 & 61$^2$\vspace*{0.15cm}\\ 
6C1256+36 & 1.128 & 21/03/99, 22/03/99 		 & 1.5-2.0 & blue & 9390   & 3100--5400 & 40$^{2}$, 22$^9$, 0$^9$\\
          &       & 21/03/99, 22/03/99 		 & 1.5-2.0 & red  & 9000   & 7000--9000 & 40$^{2}$, 22$^9$, 0$^9$\\
	  &       & 21/03/99, 22/03/99 		 & 1.5-2.0 & blue & 6240   & 3100--5400 & 115$^9$, 79$^9$\\
          &       & 21/03/99, 22/03/99 		 & 1.5-2.0 & red  & 6000   & 7250--9000 & 115$^9$, 79$^9$\vspace*{0.15cm}\\
6C1257+36 & 1.004 & 28/02/00           		 & 2.0-2.5 & blue & 11720  & 3100--5400 & 317$^2$\\
          &       & 28/02/00           		 & 2.0-2.5 & red  & 11220  & 6750--8250$^7$ & 317$^2$\vspace*{0.05cm}\\\hline 
\multicolumn{8}{l}{Notes:} \\
\multicolumn{8}{l}{[1]: The seeing conditions on 01/03/00 were extremely
poor. Some blue arm observations on this night were necessarily
excluded due}\\
\multicolumn{8}{l}{ to their very low signal--to--noise.}\\
\multicolumn{8}{l}{[2]: Slit aligned along the radio axis.} \\
\multicolumn{8}{l}{[3]: Blue arm data beyond the dichroic is included
because of the high signal--to--noise ratio of the Mg\textsc{ii} 2800 line}\\
\multicolumn{8}{l}{[4]: Slit aligned with the optical emission.} \\
\multicolumn{8}{l}{[5]: The red arm observations of 6C1011+36 on
01/03/00 were excluded due to the very poor seeing conditions.}\\ 
\multicolumn{8}{l}{[6]: The [Ne\textsc{v}]3426 line for 6C1019+39 is
very weak, and only one red arm central wavelength was used in order
to maximise the}\\
\multicolumn{8}{l}{ signal--to--noise ratio of this emission line. }\\
\multicolumn{8}{l}{[7]: Only one red arm central wavelength was used for this
object, due to the loss of the latter half of the night of 01/03/00.}\\
\multicolumn{8}{l}{[8]: The red arm exposure time for 6C1217+36 is
considerably lower than that for the blue arm, due to technical problems during}\\
\multicolumn{8}{l}{ the observations.} \\
\multicolumn{8}{l}{[9]: Slit angled in other directions to
include other objects in the field.} \\
\end{tabular}
\end{center}
\end{table*}

\begin{table*}
\scriptsize{}
\caption{Spectroscopic properties of the 6C radio galaxies. The
[O\textsc{ii}] 3727\AA\, integrated flux 
(units of $10^{-19}\rm{W\,m}^{-2}$) corresponds to the
line flux along the entire length of the slit calculated by
integrating the \oo 3727\AA\, intensities shown in
Figs.~\ref{Fig: 1}--\ref{Fig: 10}(d). All the flux ratios and flux
densities quoted in this table are corrected for galactic extinction
using the {\it E(B-V)} for the Milky Way from the NASA Extragalactic
Database (NED)
and the parameterized galactic extinction law of Howarth (1983). All
flux ratios are measured relative to \oo 3727\AA\, (value of
100) within the extracted one-dimensional spectrum.  The error on the
\oo 3727\AA\, line flux is dominated by calibration errors, estimated to be
$\lta 10$\%. Errors on the other lines and this calibration error are
added in
quadrature with the errors arising from photon statistics. There may
be a small $\lta 5$-10\% systematic offset between the red and blue
arm measurements due to different spatial extraction regions - the
extracted regions are of the same spatial extent, but may not be
centred on the same position. Galaxies may not always be completely
centred in the slit at the red and blue extremes due to differential refraction,
as the slit position angle was not necessarily aligned with the
parallactic angle. This should be minimal however, as observations
were generally taken at as low an airmass as possible, with a maximum
value typically $< 1.3$.}
\footnotesize
\begin{center} 
\begin{tabular} {lccccccccc} 
\multicolumn{2}{c}{Source}   &{6C0943}&{6C1011}&{6C1017} &{6C1019} & {6C1129} & {6C1217} & {6C1256} & {6C1257}\\
\multicolumn{2}{c}{Redshift} & {1.035} & {1.042} & {1.053} & {0.922} & {1.060} & {1.088} & {1.128} & {1.004}\\
\multicolumn{2}{c}{Radio Size (kpc)} & {92} & {444} & {65} & {67}    & {141}   & {38}    & {155}   & {336}\\ 
\multicolumn{2}{c}{Milky Way E(B-V)}& {0.018} & {0.013}& {0.008}& {0.013}& {0.030}& {0.017}&{0.014}& {0.014}\\
\multicolumn{2}{c}{Integrated [O\textsc{ii}] 3727\AA\, flux} & {5.01} &{1.24} & {6.86} & {2.43} & {5.14} & {0.51} & {2.09} & {2.24}\vspace*{0.15cm}\\
{CIV 1549}  & {Flux Ratio}            & {38.0} & {543}  & {123}  & {*}    & {29.3} & {*}    & {45.0} & {*}\\
            & {Error}                 & {9.8}  & {111}  & {24.9} & {*}    & {7.6}  & {*}    & {22.6} & {*}\\
            & {Equiv. width}          & {103}  & {90}   & {151}  & {*}    & {-}    & {*}    & {-}    & {*}\vspace*{0.05cm}\\
{He\textsc{ii} 1640} & {Flux Ratio}   & {36.8} & {197}  & {59.0} & {*}    & {16.7} & {*}    & {34.5} & {*}\\
            & {Error}                 & {8.9}  & {42.8} & {12.4} & {*}    & {4.9}  & {*}    & {17.3} & {*}\\
            & {Equiv. width}          & {97}   & {34}   & {66}   & {*}    & {39}   & {*}    & {-}    & {*}\vspace*{0.05cm}\\
{C\textsc{iii}] 1909} & {Flux Ratio}  & {29.6} & {161}  & {57.9} & {58.8} & {22.4} & {42.4} & {18.7} & {69.6}\\
            & {Error}                 & {6.1}  & {32.7} & {11.9} & {22.7} & {4.8}  & {18.2} & {9.4}  & {22.2}\\
            & {Equiv. width} 	      & {49}   & {43}   & {58}   & {43}   & {58}   & {22}   & {-}    & {25}\vspace*{0.05cm}\\
{C\textsc{ii}] 2326} & {Flux Ratio}   & {15.7} & {33.9} & {20.4} & {22.2} & {1.9}  & {35.1} & {9.1}  & {12.1}\\
            & {Error}        	      & {3.4}  & {7.3}  & {4.1}  & {4.6}  & {1.5}  & {25.6} & {4.6}  & {5.0}\\
            & {Equiv. width} 	      & {23}   & {9}    & {24}   & {65}   & {-}    & {18}   & {-}    & {14}\vspace*{0.05cm}\\
{[NeIV] 2425} & {Flux Ratio} 	      & {10.9} & {87.9} & {23.6} & {4.5}  & {16.0} & {7.8}  & {4.3}  & {23.2}\\
            & {Error}        	      & {2.4}  & {17.4} & {4.7}  & {2.9}  & {4.6}  & {7.4}  & {2.2}  & {7.1}\\
            & {Equiv. width} 	      & {17}   & {26}   & {28}   & {14}   & {58}   & {-}    & {-}    & {27}\vspace*{0.05cm}\\
{Mg\textsc{ii} 2798} & {Flux Ratio}   & {*}    & {185}  & {*}    & {-}    & {*}    & {*}    & {*}    & {*}\\
            & {Error}        	      & {*}    & {50.5} & {*}    & {-}    & {*}    & {*}    & {*}    & {*}\\
            & {Equiv. width}          & {*}    & {76}   & {*}    & {-}    & {*}    & {*}    & {*}    & {*}\vspace*{0.05cm}\\
{[NeV] 3426} & {Flux Ratio}           & {16.1} & {83.1} & {25.4} & {2.1}  & {11.6} & {5.8}  & {14.8} & {28.6}\\
            & {Error}                 & {4.7}  & {17.6} & {5.1}  & {1.9}  & {3.0}  & {4.6}  & {5.8}  & {6.4}\\
            & {Equiv. width}          & {36}   & {38}   & {45}   & {-}    & {52}   & {3}    & {26}   & {33\vspace*{0.05cm}}\\
{[O\textsc{ii}] 3727} & {Flux Ratio}  & {100}  & {100}  & {100}  & {100}  & {100}  & {100}  & {100}  & {100}\\
            & {Equiv. width}          & {216}  & {41}   & {195}  & {58}   & {226}  & {59}   & {147}  & {83}\vspace*{0.05cm}\\
{[Ne\textsc{iii}] 3869} & {Flux Ratio}& {35.5} & {98.4} & {32.5} & {17.3} & {30.0} & {5.8}  & {17.7} & {22.8}\\
            & {Error}                 & {8.3}  & {20.7} & {6.6}  & {4.7}  & {6.0}  & {5.4}  & {4.9}  & {4.9}\\
            & {Equiv. width}          & {94}   & {39}   & {69}   & {11}   & {56}   & {4}    & {34}   & {25}\vspace*{0.05cm}\\
{H$\zeta$ 3889}   & {Flux Ratio}      & {-}    & {21.0} & {7.3}  & {-}    & {-}    & {-}    & {12.4} & {8.5}\\
            & {Error}        	      & {-}    & {6.7}  & {1.5}  & {-}    & {-}    & {-}    & {3.7}  & {2.8}\\
            & {Equiv. width}	      & {-}    & {8}    & {16}   & {-}    & {-}    & {-}    & {24}   & {10}\vspace*{0.05cm}\\
{H$\epsilon$ + [Ne\textsc{iii}] 3967}&{Flux Ratio}&{5.9}&{33.9}&{12.5}&{-}& {9.7}  & {13.7} & {-}    & {13.4}\\
            & {Error}                 & {2.7}  & {9.5}  & {2.5}  & {-}    & {2.5}  & {6.4}  & {-}    & {13.4}\\
            & {Equiv. width}          & {9}    & {13}   & {29}   & {-}    & {17}   & {6}    & {-}    & {12}\vspace*{0.05cm}\\
{H$\delta$ 4102} & {Flux Ratio}       & {-}    & {10.5} & {19.0} & {-}    & {-}    & {-}    & {-}    & {-}\\
            & {Error}                 & {-}    & {9.1}  & {4.0}  & {-}    & {-}    & {-}    & {-}    & {-}\\
            & {Equiv. width}          & {-}    & {-}    & {53}   & {-}    & {-}    & {-}    & {-}    & {-}\vspace*{0.05cm}\\
{H$\gamma$ 4340} & {Flux Ratio}       & {-}    & {49.2} & {16.3} & {*}    & {*}    & {*}    & {*}    & {*}\\
            & {Error}                 & {-}    & {14.3} & {3.2}  & {*}    & {*}    & {*}    & {*}    & {*}\\
            & {Equiv. width}          & {-}    & {11}   & {47}   & {*}    & {*}    & {*}    & {*}    & {*}\vspace*{0.15cm}\\
{Mean Flux Density} & {2100-2300\AA} & {1.45} & {2.41} & {2.60} & {0.93} & {0.72} & {1.92} & {0.01} & {1.73}\\
{Error} &                            & {0.21} & {0.30} & {0.32} & {0.32} & {0.13} & {0.47} & {0.04} & {0.49}\vspace*{0.05cm}\\
{Mean Flux Density} & {2450-2700\AA} & {1.17} & {-}    & {-}    & {0.96} & {-}    & {-}    & {-}    & {1.01}\\
{Error} &                            & {0.23} & {-}    & {-}    & {0.32} & {-}    & {-}    & {-}    & {0.37}\vspace*{0.05cm}\\
{Mean Flux Density} & {3500-3700\AA} & {1.01} & {1.22} & {1.63} & {2.63} & {0.74} & {1.01} & {0.52} & {1.67}\\
{Error} &                            & {0.16} & {0.20} & {0.23} & {0.33} & {0.15} & {0.27} & {0.12} & {0.25}\vspace*{0.05cm}\\
{Mean Flux Density} & {4050-4250\AA} & {1.09} & {2.11} & {0.68} & {-}    & {-}& {-}    & {0.87} & {1.91$^\dag$}\\
{Error} &                            & {0.31} & {0.57} & {0.46} & {-}    & {-}& {-}    & {0.39} & {0.53$^\dag$}\\
\normalsize
\end{tabular}
\end{center}
\end{table*}

\addtocounter{table}{-1}
\begin{table*}
\scriptsize{}
\caption{\textbf{-- Continued.}   
Equivalent widths are measured in the rest frame of the
galaxy. A single dash is used to represent emission lines of such low
flux as to be unobservable, and asterisks are used to represent
emission lines outside the wavelength range covered by our spectra.
Mean continuum flux densities for line-free wavelength ranges (or
as much of the range as possible provided at least 100\AA\, are covered
in the spectrum) are measured from the extracted one-dimensional
spectrum; the (weak) H$\delta$ emission line was subtracted from the
spectrum before calculating the mean continuum level for the
wavelength range 4050--4250\AA. Values are in units of
$10^{-21}\rm{W\,m}^{-2}\AA^{-1}$; the uncertainty given
is the error on the mean value in that wavelength region.
\dag\,\, indicates that a shorter wavelength range has been used in
measuring the flux density. }
\begin{center} 
\begin{tabular} {lcc} 
\end{tabular}
\end{center}
\end{table*}

The layout of the paper is as follows.  Section 2 gives details of the
sample selection, the observations, and the data reduction
techniques. In Section 3, the results of the spectroscopic
observations are presented, including a two-dimensional analysis of
the \oo 3727\AA\, emission line. Section 4 describes an analysis of
the ionization and kinematical properties of the emission line gas and
compares these with the results of the more powerful 3CR galaxies at
the same redshift.  Our results are discussed in section 5, and a
summary of our conclusions is given in section 6.  

Throughout the paper, values for the cosmological parameters of
$\Omega_0=0.3$, $\Omega_{\Lambda}=0.7$ and $H_{0} = 65
\rm{km\,s}^{-1}$Mpc$^{-1}$ are assumed.   

\section{Observations}

\subsection{Sample Selection and Observational Procedures}

The observed radio galaxies were selected from the 6CER
sample\footnote{Up--to--date information for the revised
6CE sample can be found at: http://www-astro.physics.ox.ac.uk/$\sim$sr/6ce.html.}   
(Rawlings, Eales \& Lacy 2001), a revised version of the sample
originally defined by Eales (1985). This revised sample is complete,
and consists of 59 radio sources with flux densities in the range
2.0\,Jy\,$<\,S_{151}\,<$\,3.93\,Jy at 151\,MHz which lie in the region
of sky $08^{h}20^{m}<RA<13^{h}01^{m}$ and $34^{\circ}<\delta<40^{\circ}$.
Radio, optical and infrared studies have been carried out for a complete
subsample of 11 radio galaxies, selected from this sample within the
redshift range $0.85<z<1.5$ (Best {\it et al} 1999, Inskip {\it et al}
in prep). The spectroscopic study of these galaxies 
was restricted to eight objects with $z<1.2$, in order to
include only those galaxies for which the observed wavelength of the
4340\AA\, H$\delta$ line would be observed at wavelengths  $\lambda
\lta 9100$\AA, and thus potentially observable.  The
three excluded galaxies all have redshifts 
$z>1.37$.  This subsample of 6C galaxies is well matched to the nearly
complete subsample of 14 3CR 
galaxies studied by Best {\it et al.} (2000a), which lies within the
same redshift range.  

The observations were carried out on 1999 March 20-22 and 2000 February
28-March 02, using the dual-beam ISIS spectrograph (Carter {\it et al} 1994) on the William
Herschel Telescope (WHT). To enable the highest throughput at short
wavelengths, the 5400\AA\, dichroic was used. In the blue arm, the
R158B grating was used in conjunction with the EEV12 CCD. This low
dispersion grating provided a spatial scale of 0.4 arcsec per pixel, a
spectral scale of 2.7\AA\, per pixel, a spectral resolution of
approximately 19\AA and a wide spectral coverage from 3000\AA\, to the
dichroic at about 5400\AA.  With this arrangement, we were able to 
measure accurately the strengths of emission lines at short wavelengths. 
In the red arm of the spectrograph, the R316R grating was used in
conjunction with the TEK4 CCD. This provided a spatial scale of 0.36
arcsec per pixel, a spectral scale of 1.49\AA\, per pixel and spectral
resolution of 
approximately 5\AA. The observed wavelength coverage in a single
grating position is 1500\AA. Two different central wavelengths were generally
used; the first value selected gave a restframe  wavelength coverage from
[Ne\textsc{v}]3426 out to approximately 4000\AA, and the second
typically covered the spectrum from 3600\AA\, to H$\gamma$, or
H$\delta$ for the higher redshift sources. The spectral range provided
by each selected central wavelength was chosen to include the
3727\AA\, \oo\, line, and also enable measurement of either the \nrat line
ratio or the Balmer line fluxes of the source in question.
Full details of the observations are given in Table 1.

\subsection{Observations and data reduction}

Long-slit spectra were obtained for all eight galaxies, with total
integration times varying between 2 to 5.5 hours, averaging 3.5 hours 
in each arm. Red arm exposures were typically split into individual
exposures of 1500s in length to aid the removal of cosmic rays.
Individual blue arm exposures were taken for twice as long in order to
reduce the fractional read-noise contribution.  These
exposures were sky background limited, due to the binning of the blue
arm CCD and the longer exposure times.  In most cases, the slit was
oriented along the radio axis of the source. For 6C1256+36, exposures
at several different sky PAs were taken, in order to include other
objects in the field suspected of being cluster members around the
radio galaxy (to be discussed elsewhere).  Two different PAs were used
for 6C1011+36 in order to position the slit along the axes of both the
radio jets and the extended optical emission.  Details of the
observations are included in full in Table 1. 
 
The seeing varied from 0.7 to 2.0 arcsec during the 1999 March
observations. For the 2000 observations, the seeing was more
changeable.  The seeing on the first night was between 2 to 2.6 arcsec,
but decreased to $\approx$1.8 arcsec throughout night 2. The seeing on the
final night increased steadily from 1.9 arcsec up to non--photometric 
conditions of greater than 3 arcsec seeing, at which point further
observations became unusable. 

Standard packages within the NOAO \textsc{iraf} reduction software
were used to reduce the raw data. Corrections were made for overscan
bias subtraction, and the data were then flat-fielded using internal
calibration lamp observations.  The flat field images were made using
the same instrumental set-up as the observations for each source, in
order to minimize fringing effects.  As red arm images of the galaxies
were taken at one or more different observed wavelength ranges for
each object, several flat fields had to be taken for each galaxy, and
the flat field exposures were interspersed between the observations
on-source. Taking care not to subtract any of the extended line emission,
the sky background was removed. 

All observations of a galaxy taken using the same central wavelength
and slit position angle were combined as 2-d spectra.  For both
6C0943+39 and 6C1011+36, the slit position angles of different
observations differ by $5^\circ$.  With such a small difference in
position angle, essentially the same regions of the galaxies are
sampled by the observations, and so the
2-d spectra of each of these sources were combined at this stage.
For 6C1256+36, five different slit position angles were used.
Examination of the 2-d structure of the emission lines showed no
apparent variation for observations with slit position angles within
$\sim 40^\circ$ of each other. The 2-d spectra of this source have
therefore been combined in two groups, with slit position angles of
$0^{\circ}$, $22^{\circ}$ \& $40^{\circ}$ and $79^{\circ}$ \&
$115^{\circ}$ respectively.   As the emission line region of this
source is $< 3^{\prime\prime}$ in extent, the observations combined at
this stage essentially sample the same region of the galaxy.  Sources
for which observations had been made at several different central
wavelengths, or very different slit PAs in the case of 6C1256+36, could not be
fully combined at this stage. One--dimensional spectra were extracted
from the central 4\arcsec\, of the combined (or partially combined) data, and wavelength
calibrated using observations of the internal CuNe and CuAr arc lamps,
which removed the large scale non-linearities.  A slight linear shift
of wavelength was sometimes also required, and was determined from
unblended sky lines (Osterbrock and Martel 1992).  The 1-d spectra of
sources taken with more than one central wavelength and/or several
slit PAs were fully combined at this point.   Observations of the
spectrophotometric standard stars  G191b2b, HZ44, G99-37 and BD+26 26,
observed in the same instrumental set-up as the observations of the
galaxies, were used to provide accurate flux calibration of the final
1-d spectra.   A comparison of our spectra with the line luminosities
of Rawlings, Eales and Lacy (2001) suggests that use of narrower slits
coupled with occasional poor seeing may have led to some slit losses,
particularly for the sources with more extensive emission line
regions. To create 2--dimensional spectra of the \oo\, line for the
analysis of the kinematic properties of the emission line gas,
observations of a particular galaxy were combined where these were
taken at sufficiently  close slit PAs, and the appearance of the \oo\,
emission lines were indistinguishable. 
\section{Results}

\subsection{Fitting the spectroscopic data}

\begin{figure*}
\vspace{7.8 in}
\begin{center}
\includegraphics{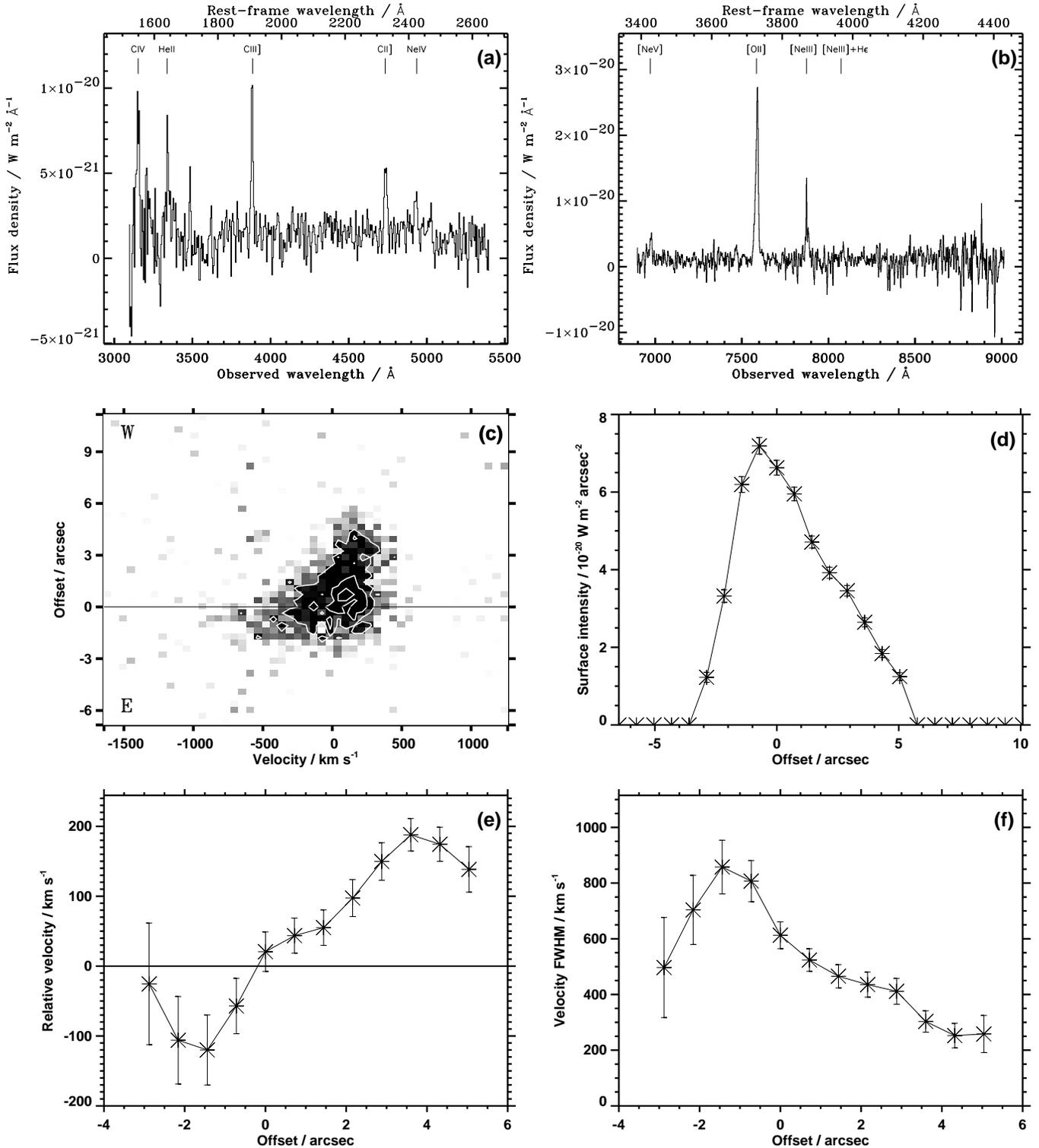}
\end{center}
\caption{Spectroscopic data for 6C0943+39. (a -- upper left) One
dimensional spectrum extracted from the blue arm. Emission lines are
labelled in both this and (b -- upper right) the one
dimensional spectrum extracted from the red arm. (c -- middle
left) Two-dimensional extracted image of the [O\textsc{ii}]3727\AA\, emission
line. Offset is measured in both `positive' and `negative' sky
directions from the continuum centroid as defined by the labelling in
the left--hand corners of this figure. Contour levels are for the
following percentages of the maximum surface brightness level: 100\%,
75\%, 50\%.   (d -- middle right) Surface
brightness of the [O\textsc{ii}]3727\AA\, emission as a function of position along
the slit. (e --lower left) Relative velocity of the fitted Gaussian
peak as a function of slit position. (f -- lower right) The variation
of the fitted Gaussian profile FWHM as a function of slit position. 
\label{Fig: 1}}
\end{figure*}

The one dimensional spectra for both red and blue arms are displayed
in panels (a) and (b) of Figs.~\ref{Fig: 1}-2, Figs.~\ref{Fig: 4}-8
and Fig.~\ref{Fig: 10}. The fluxes of various emission lines relative
to \oo 3727\AA\, are tabulated in Table 2, as well as their equivalent
widths and the mean continuum flux density in various regions of the
spectrum. The flux ratios and flux densities tabulated in Table 2 have been
corrected for galactic extinction, using values for the H\textsc{i} column density
of the Milky Way taken from the NASA Extragalactic Database (NED), and
the parameterized galactic extinction law of Howarth (1983).  The
redshifts obtained from the spectra of the eight sources agree with
those of Rawlings, Eales \& Lacy (2001), confirming less-certain
single-line redshifts in four cases.

\begin{figure*}
\vspace{8.2 in}
\begin{center}
\includegraphics{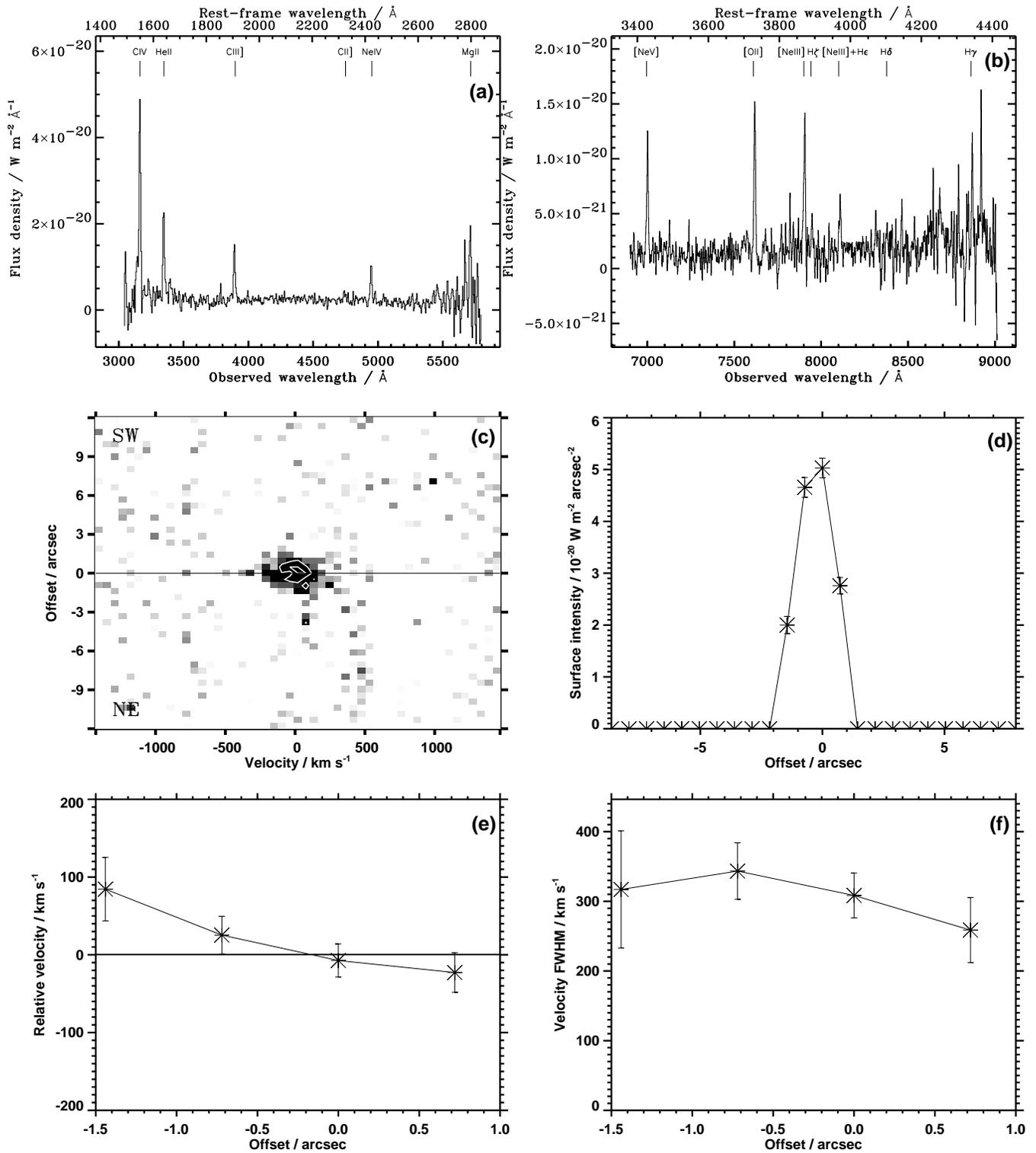}
\end{center}
\caption{Spectroscopic data for 6C1011+36 with the slit aligned along
the radio axis at an angle of 47$^{\circ}$. Contour levels in (c) are
at 75\% and 100\% of the maximum surface brightness level. Offset
directions are as labelled in frame (c).  Other details as in Fig. 1.
\label{Fig: 2}}
\end{figure*}

\begin{figure*}
\vspace{5.0 in}
\begin{center}
\includegraphics{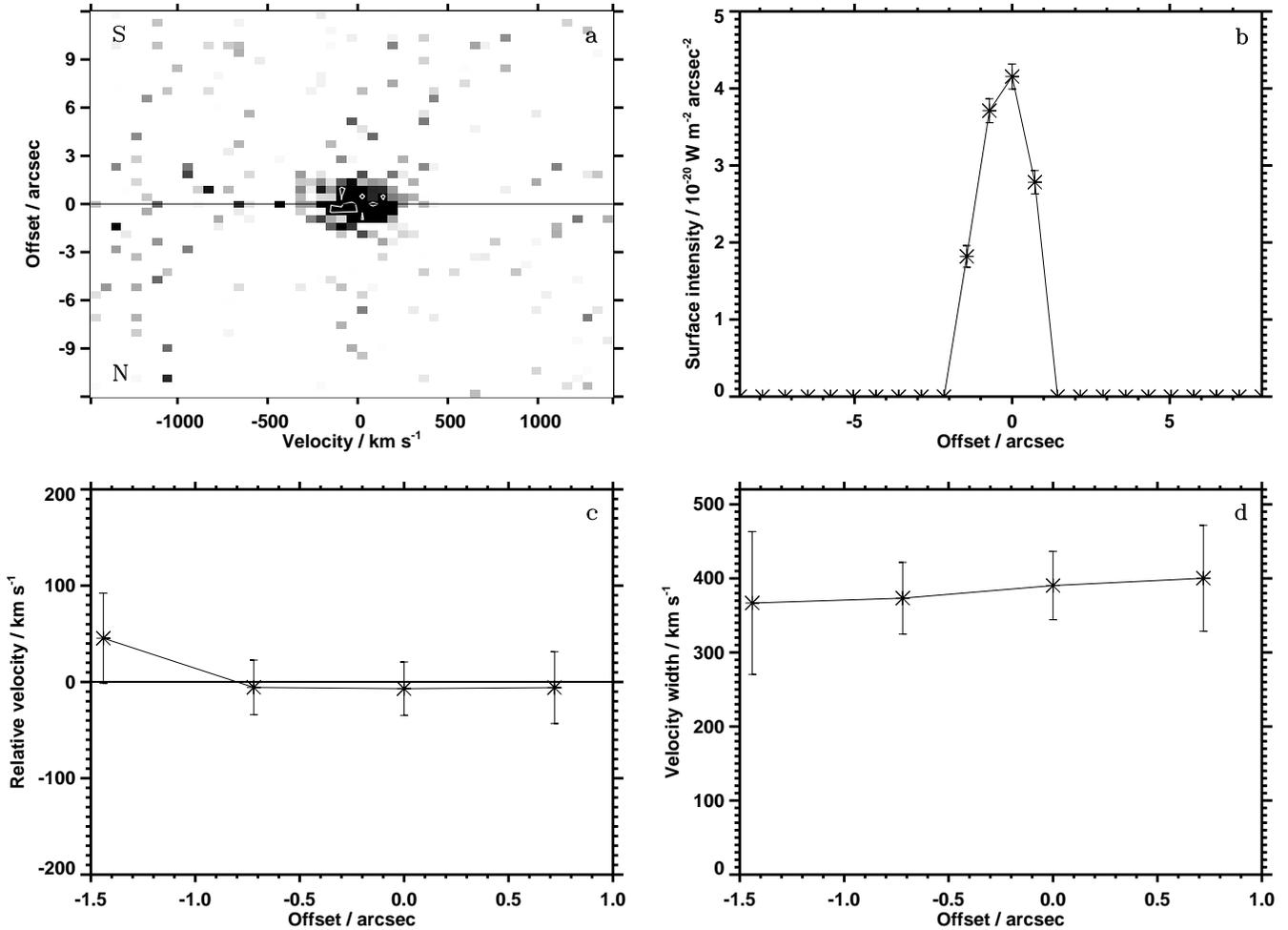}
\end{center}
\caption{Spectroscopic data for 6C1011+36, at a sky PA of
345$^{\circ}$ (aligned with the optical 
emission, at $\sim$60$^{\circ}$ to the radio axis).
Figs (a) to (d) match captions (c) to (f) in Fig.~\ref{Fig: 1}. Offset
directions are as labelled in frame (a). Contour levels in (a) are
at 90\% of the maximum surface brightness level. 
\label{Fig: 3}}
\end{figure*}

\begin{figure*}
\vspace{8.2 in}
\begin{center}
\includegraphics{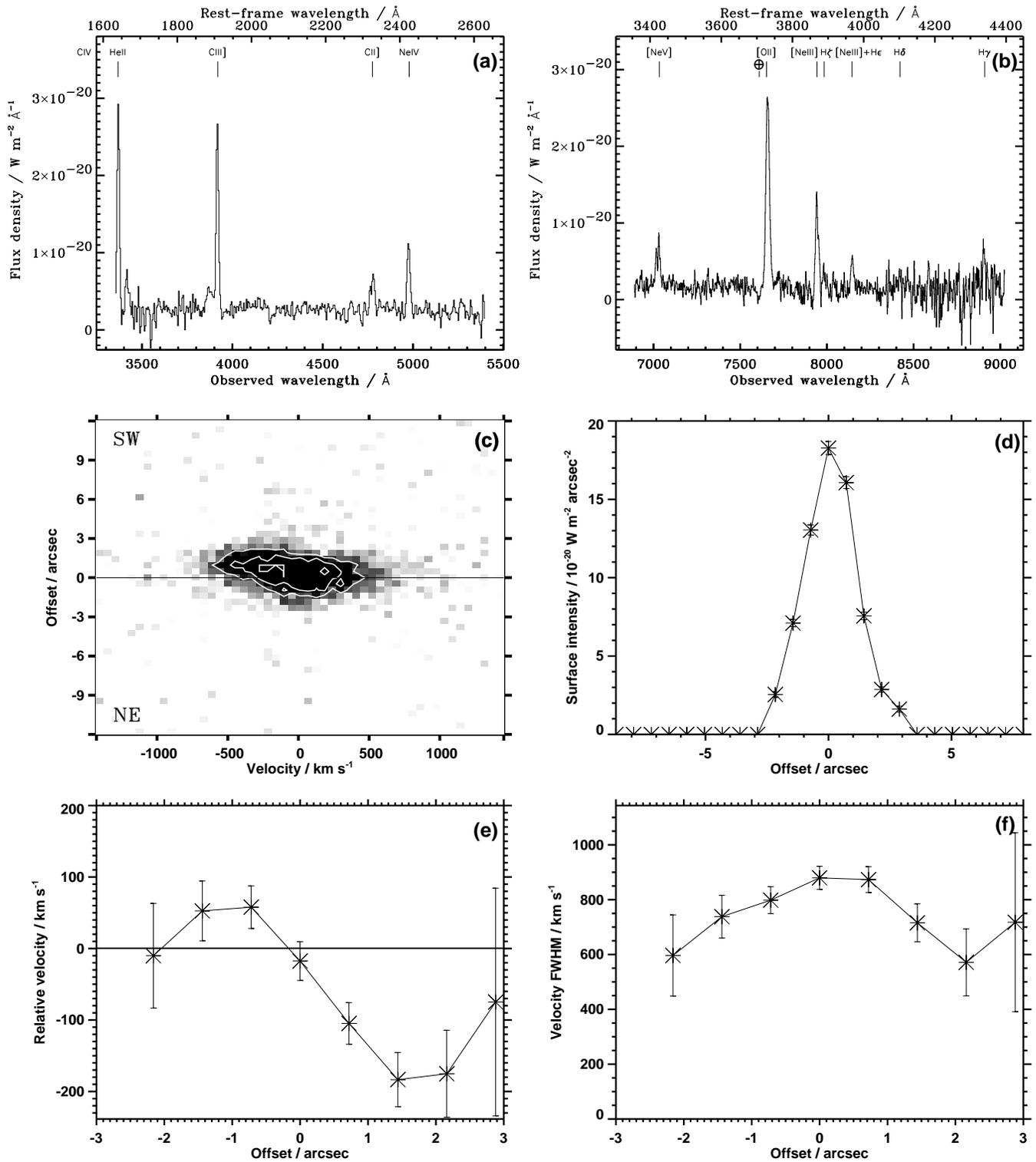}
\end{center}
\caption{Spectroscopic data for 6C1017+37. Offset
directions are as labelled in frame (c).  Contour levels  in frame
(c) are for 50\%, 75\% and 100\% of the maximum surface brightness
level. Other details as in Fig. 1. 
\label{Fig: 4}}
\end{figure*}

\begin{figure*}
\vspace{8.2 in}
\begin{center}
\includegraphics{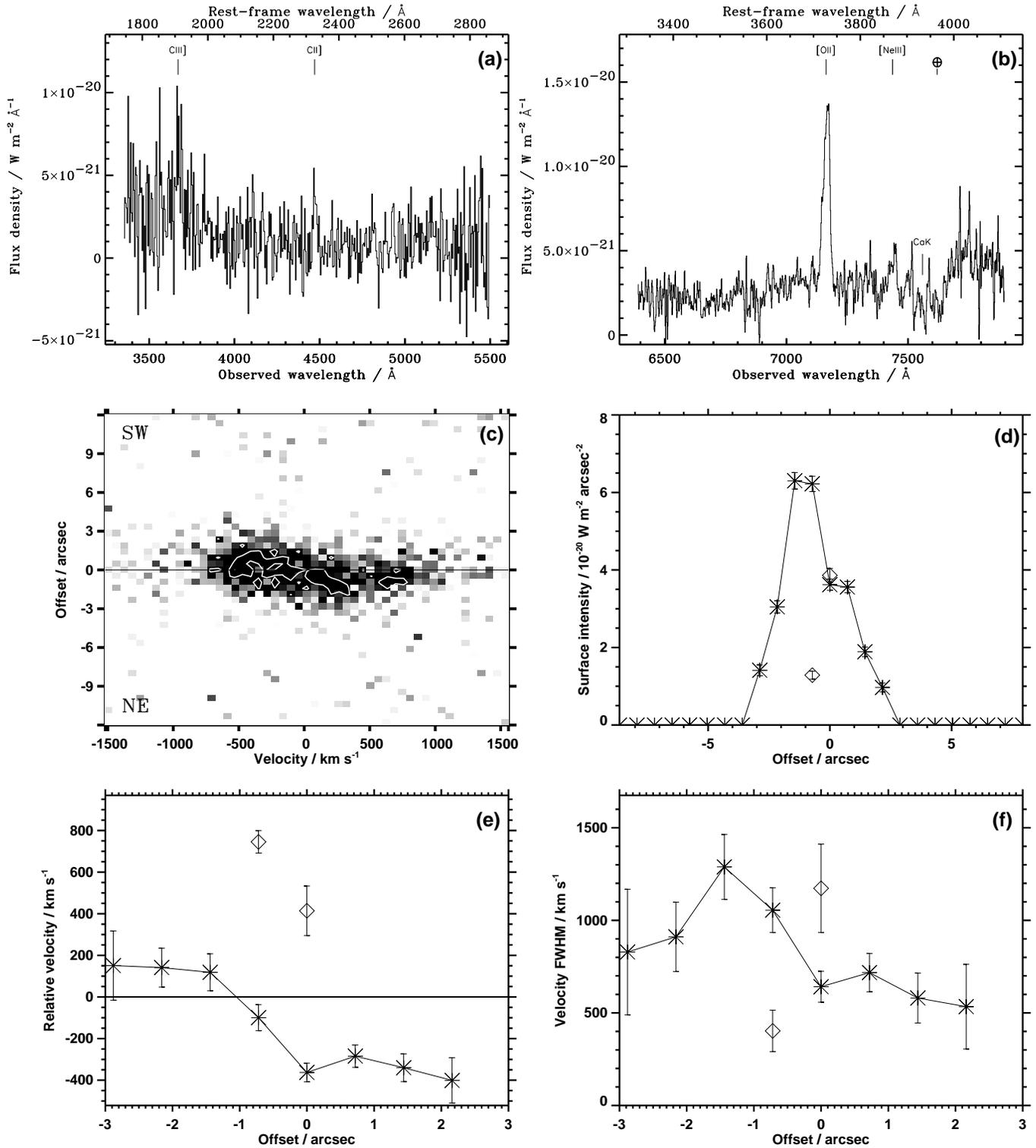}
\end{center}
\caption{Spectroscopic data for 6C1019+39. Contour levels in (c) are
at 75\% and 100\% of the maximum surface brightness level. 
For this object a second Gaussian was fitted to the data which is
plotted in the figures using open diamonds.  Offset
directions are as labelled in frame (c).  Other details as in Fig. 1.
\label{Fig: 5}}
\end{figure*}

\begin{figure*}
\vspace{8.2 in}
\begin{center}
\includegraphics{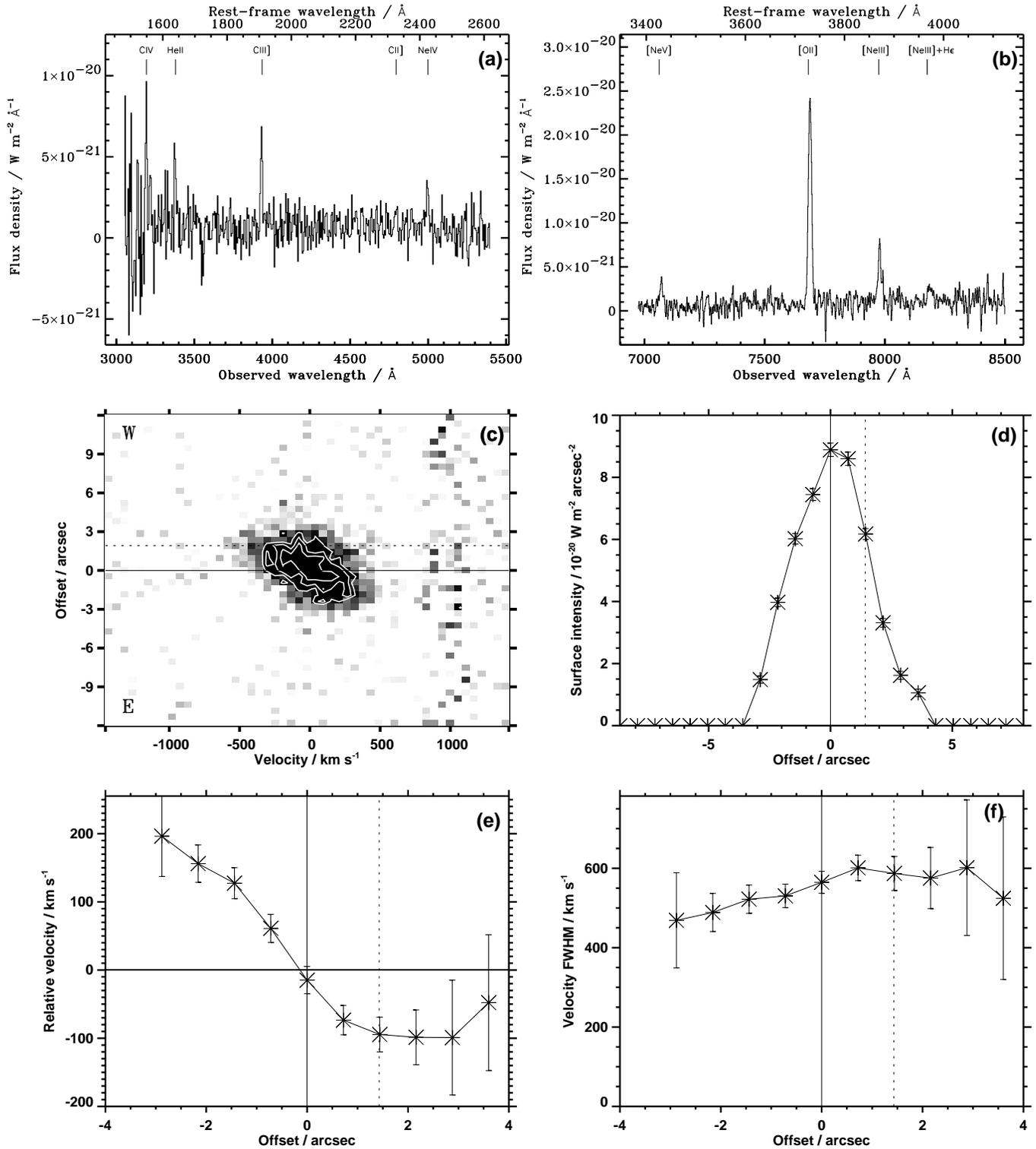}
\end{center}
\caption{Spectroscopic data for 6C1129+37. Contour levels in (c) are
at 60\%, 75\%, and 90\% of the maximum surface brightness level. Offset
directions are as labelled in frame (c). The
solid line at offset zero represents the position of the eastern
galaxy, tentatively identified as the host of the radio source. The
position of the western galaxy is represented by the dotted line. 
Other details as in Fig. 1.
\label{Fig: 6}}
\end{figure*}

\begin{figure*}
\vspace{8.2 in}
\begin{center}
\includegraphics{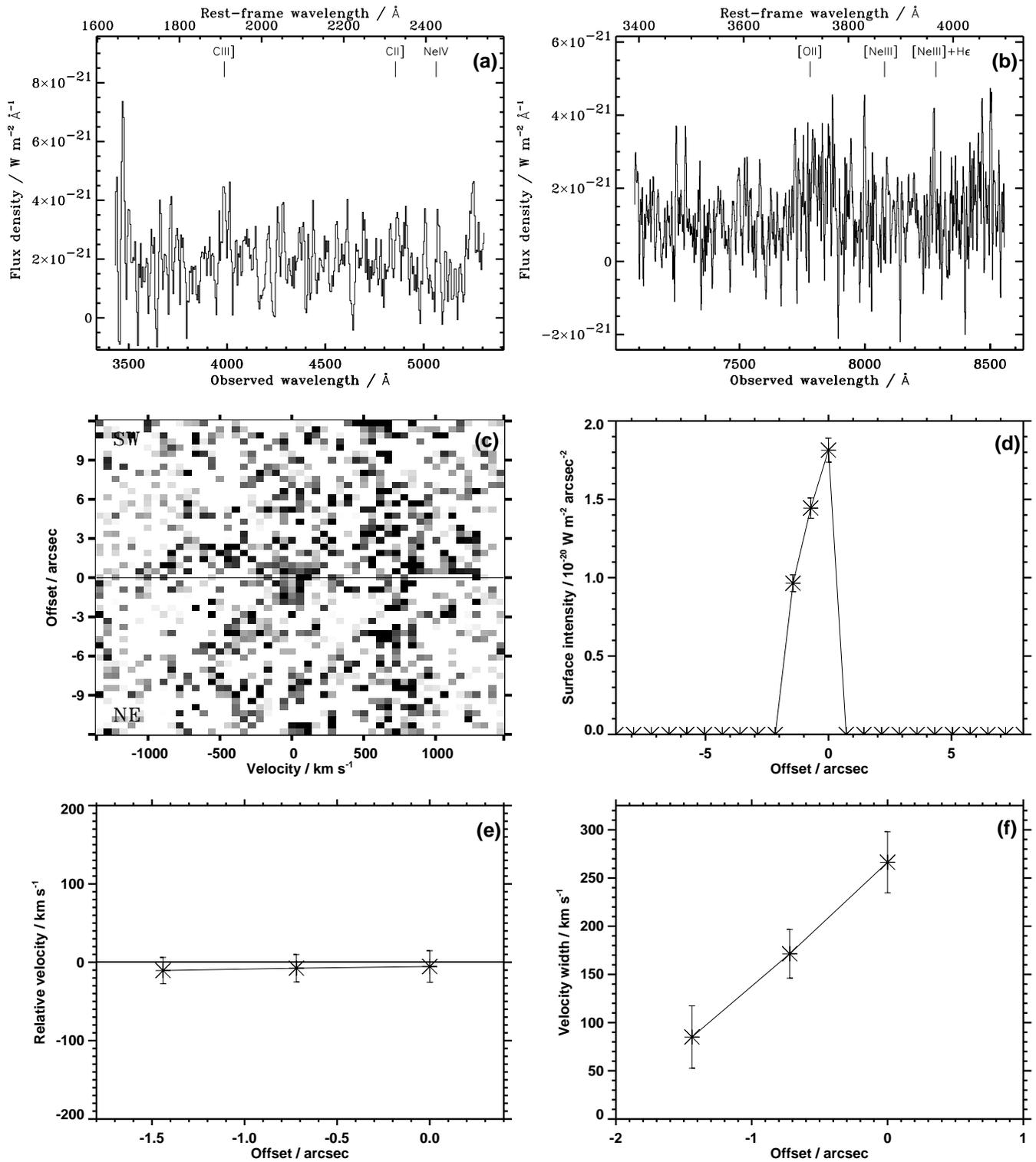}
\end{center}
\caption{Spectroscopic data for 6C1217+36. Offset
directions are as labelled in frame (c).  Other details as in Fig. 1. The
\oo\, emission line in this source is particularly faint.
\label{Fig: 7}}
\end{figure*}

\begin{figure*}
\vspace{8.2 in}
\begin{center}
\includegraphics{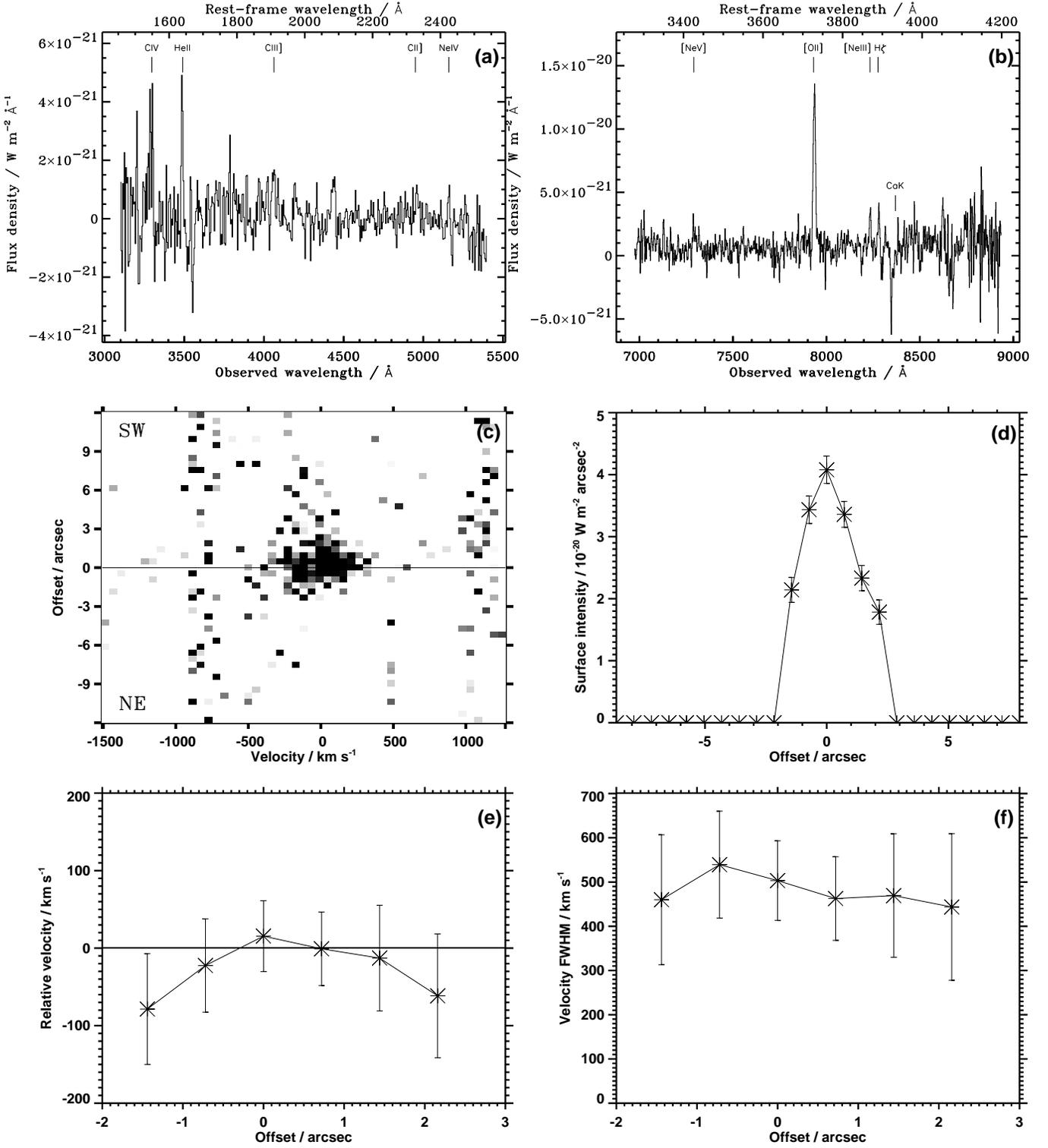}
\end{center}
\caption{Spectroscopic data for 6C1256+36, with the slit aligned along
the radio axis (40$^\circ$). The two dimensional \oo\, spectrum combines
data from the relatively close sky PAs of 40$^\circ$, 0$^\circ$ and
22$^\circ$.  Offset
directions are as labelled in frame (c).  Other details as in Fig. 1. 
\label{Fig: 8}}
\end{figure*}

\begin{figure*}
\vspace{5.0 in}
\begin{center}
\includegraphics{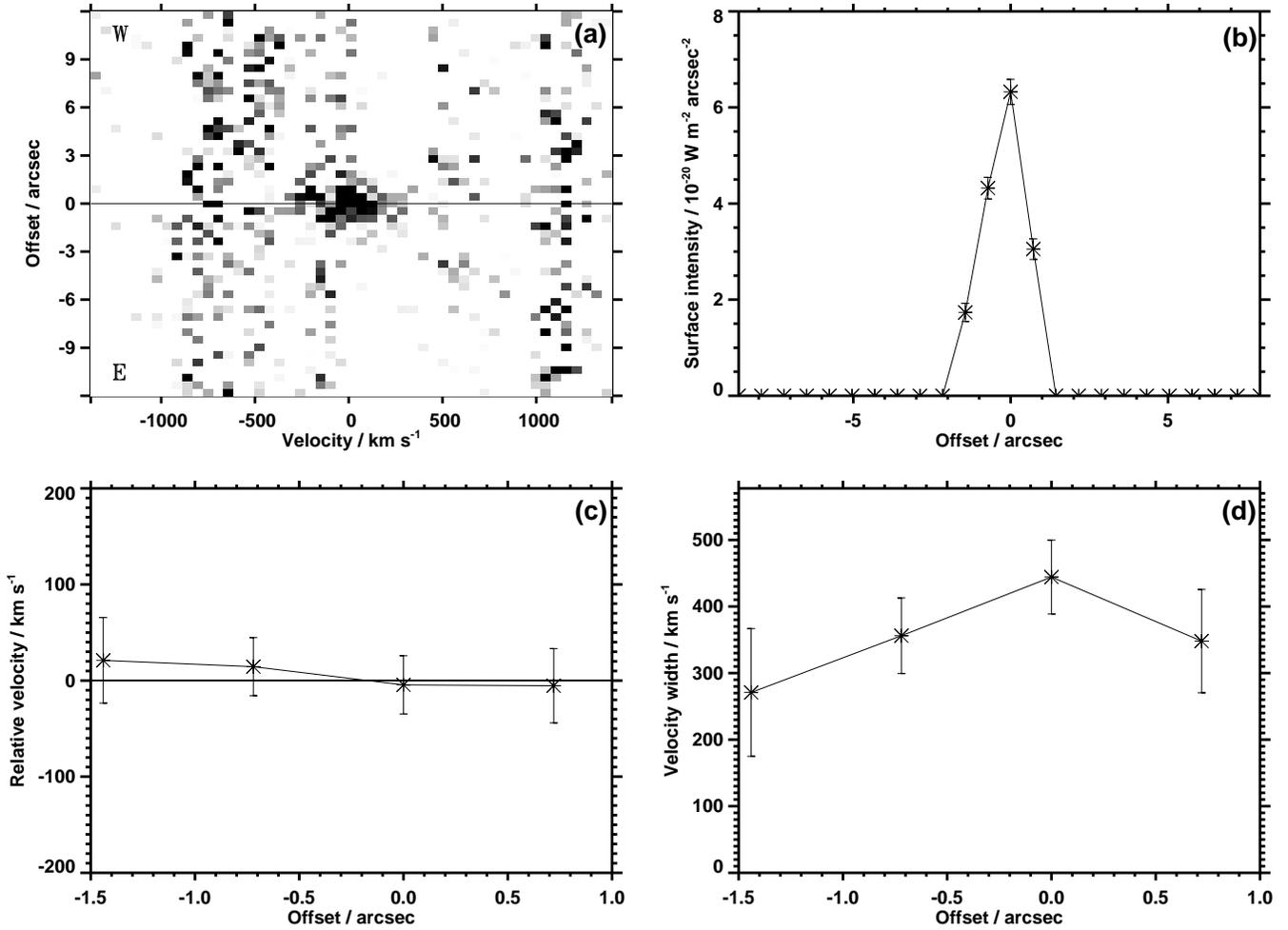}
\end{center}
\caption{Spectroscopic data for 6C1256+36, using data combined from
observations at sky PAs of 79$^{\circ}$ and 115$^{\circ}$. Offset
directions are as labelled in frame (a).  Other details as in Fig. 3.
\label{Fig: 9}}
\end{figure*}

\begin{figure*}
\vspace{8.2 in}
\begin{center}
\includegraphics{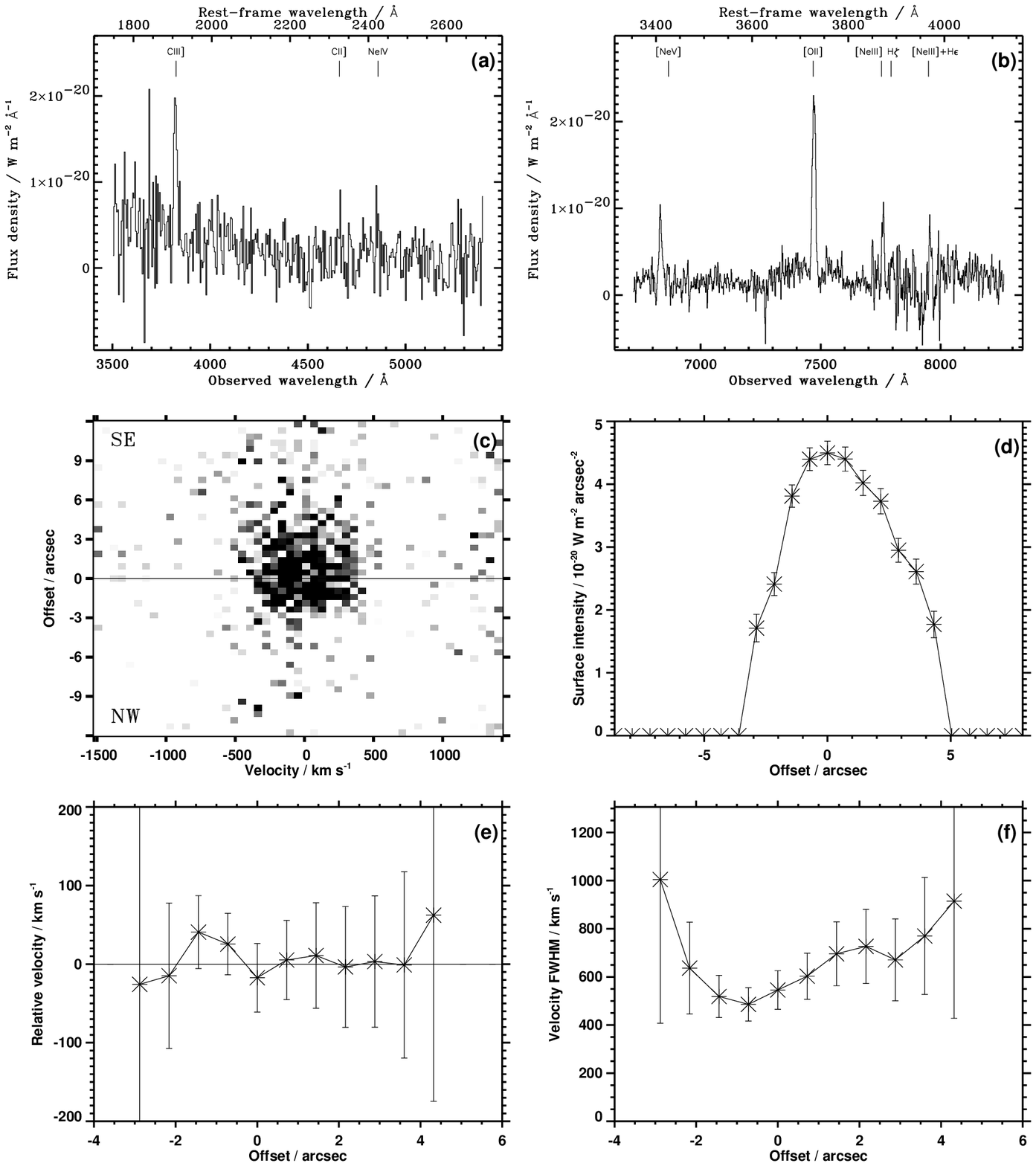}
\end{center}
\caption{Spectroscopic data for 6C1257+36. Offset
directions are as labelled in frame (c).  Other details as in Fig. 1.
\label{Fig: 10}}
\end{figure*}

A two--dimensional region around the \oo 3727\AA\, emission line was
extracted in order to study the velocity structure of the gas [Figs.~\ref{Fig:
1}-2(c), Fig.~\ref{Fig: 3}(a), Figs.~\ref{Fig: 4}-8(c),
Fig.~\ref{Fig: 9}(a) and Fig.~\ref{Fig: 10}(c)]. From these, a
sequence of one dimensional spectra were 
extracted along the slit direction, stepped every $0.72$ arcsec
(2 pixels), with widths of $1.44$ arcsec (4 pixels). The
extracted spectra were then analysed following the procedure described
in Best {\it et al.} (2000a), which is summarised below. 

After allowing for continuum subtraction, the data were fitted by
successive Gaussian components, which were only accepted if the
signal-to-noise ratio was greater than five and the FWHM larger than
the instrumental resolution.  The combination of Gaussians providing
the best fit (that with the lowest reduced $\chi^{2}$) to the
extracted spectra determined the maximum number of velocity components
which could realistically be fitted to the data. 

This approach allows us to search for high
velocity components in the emission line gas, or other structures
incompatible with a fit to a single velocity component. The integrated
emission line flux, the velocity relative to that at the centre of
the galaxy, and the emission line FWHM were determined for each
Gaussian fit. In order to calculate the emission line FWHM, it was
necessary to deconvolve it by subtracting in quadrature the
instrumental FWHM, as determined from unblended sky lines. This procedure
assumes that the line emission illuminates the slit in a similar
manner to the background sky emission, which is an adequate
approximation for narrow slits and imperfect seeing conditions. 
Errors were also calculated for these three parameters. 

This method is optimised for determining the variation of the kinematic
properties of the galaxies with position very accurately. However, the
flux levels in each extracted one--dimensional spectrum as a
percentage of the total \oo\, emission line flux of the galaxy are
fairly low. This, and the low S/N levels in the spectroscopic data,
makes identification of any broad emission line structures, such as
those found by Sol\'{o}rzano-I\~{n}arrea {\it et al.} (2001), much
more difficult.  A single Gaussian fit was not always ideal for our
data, with the profiles of some sources exhibiting weak wings on
either the red or blue side of the line. These may simply indicate
that the intrinsic line-shape is non-Gaussian or possibly could be 
associated with a weaker emission component at a different velocity
which was too faint to be individually identified, or a faint broad
line component.  By extracting a lower spatial resolution spectrum
(typically 7--9 pixels instead of 4) from the two-dimensional \oo\,
line data, the emission line flux is increased, improving the
likelihood of identifying an underlying broad component. However, by
extracting the data over a larger spatial extent of the emission line
region, any variations in the kinematics with position introduce
ambiguities into the extracted one--dimensional spectrum. A search for
broad components in the emission lines of the 6C galaxies and the 3CR
sample of Best {\it et al} (2000a) was carried out, but gave
inconclusive results (see Appendix A). 

The results of fitting narrow lines to the two--dimensional \oo\,
emission line 
are illustrated in panels (d) to (f) of Figs.~\ref{Fig: 1}-2,
Figs.~\ref{Fig: 4}-8 and Fig.~\ref{Fig: 10}, and by panels (b) to
(d) of Figs.~\ref{Fig: 3} and 9, as detailed in the figure
captions.  Noteworthy features for all objects are
outlined below.

{\bf 6C0943+39} (Fig.~\ref{Fig: 1}) has the most
irregular two dimensional \oo 3727\AA\, emission line profile of all eight
6C galaxies studied. The emission line region is fairly large with a
spatial extent of $\sim70\,\rm{kpc}$, and the FWHM varies greatly with
position, as can be seen from its asymmetric shape. The FWHM of the
\oo\, line  at 860 kms$^{-1}$ is greater than average for the sample.
6C0943+39 has a high \oo\, luminosity and the  strengths of some of
the other emission lines from this source are amongst the strongest in
the sample, although the Balmer lines in the spectra are so weak that
they can not be detected to the 3-sigma limit of $2.65 \times
10^{-20}\rm{W\,m^{-2}}$ (assuming a velocity width of
$300\,\rm{kms}^{-1}$).  

{\bf 6C1011+36} was observed at two different sky PAs: with the slit
aligned along the radio axis (Fig.~\ref{Fig: 2}) and at
$\approx60^{\circ}$ to this, along the axis of the extended optical
emission (Fig.~\ref{Fig: 3}). The two dimensional structure of the 
\oo\, line appears similar in both orientations of the slit, and is
particularly 
compact. 6C1011+36 has the joint smallest physical size for the 
emission line region, and the lowest maximum FWHM at 400
kms$^{-1}$. Its profile is very smooth, and probably consistent with rotation.
One interesting point is that the Balmer lines of this source are
relatively luminous, although it has the second lowest integrated
\oo\, flux of the 
sample.

{\bf 6C1017+37} (Fig.~\ref{Fig: 4}) is one of the smaller
radio sources and possesses a strong \oo 3727\AA\, emission line. Its profile is
relatively smooth, a feature more characteristic of larger radio
sources (Best {\it et al} 2000a, b).
The maximum FWHM and velocity range are similar to those of other
small sources, but its emission line ratios are closer to those of the
larger ones. This apparent dichotomy will be discussed in more detail
later.
 
{\bf 6C1019+39} (Fig.~\ref{Fig: 5}) has fairly weak carbon emission
lines compared to other galaxies in this sample, but has perhaps the
most prominent 4000\AA\, break, although the strength of this break
may be influenced by the presence of the 7600\AA\, sky absorption
feature, for which no correction has been made.  Balmer lines are not
identified to a 3-sigma flux density limit of $3.5 \times
10^{-20}\rm{W\,m^{-2}}$ (assuming a velocity width of
$300\,\rm{kms}^{-1}$).   The two dimensional structure of
the \oo 3727\AA\, emission line is also unique in the irregularity of
its structure.  It appears that 6C1019+39 is fairly likely to possess
more than one velocity component in the emission line gas, including a
high velocity component at 700$\rm{km}\,\rm{s}^{-1}$ which seems to be
spatially close to the centre of the galaxy and appears to have a
fairly narrow FWHM. This high velocity component is similar to others
observed at high and low redshifts, e.g. 3C265 and 3C405 respectively
(Tadhunter 1991).  Such high velocity components may provide direct
evidence for shock acceleration of the emission line gas due to
interactions with the radio jet (Sol\'{o}rzano-I\~{n}arrea {\it et al}
2001).

{\bf 6C1129+37} (Fig.~\ref{Fig: 6}) does not exhibit any observable
Balmer lines, but other emission lines appear reasonably prominently
in these spectra.  Optical and infrared imaging of this source (Inskip
{\it et al}, in prep) shows a fair amount of diffuse aligned emission,
and two elliptical galaxies, perhaps interacting. The east galaxy is
roughly 0.1 magnitudes brighter in the $K$-band than that to the west,
and whilst the radio data shows no central core (Best {\it et al}
1999) we tentatively identify the radio source with the eastern galaxy
of the two, which appears to lie more directly between the two lobes
of radio emission than the western galaxy does.  The two
dimensional image of the \oo\, line has been centred (offset=0 on
Fig.~\ref{Fig: 6}) on the east galaxy; the position of the eastern
galaxy is identified by the 
solid line, and is coincident with the peak in the \oo\, emission. The
position of the western galaxy is identified by the dotted line.
Further detailed discussion of this source is given in the paper
presenting the imaging data.  The velocity profile of the \oo\, line
for this source appears flat near the companion galaxy.  The mean
velocity of the  
extended emission line region gas surrounding the radio galaxy varies
smoothly with spatial 
position and is consistent with either rotation or outflow.

\begin{figure*}
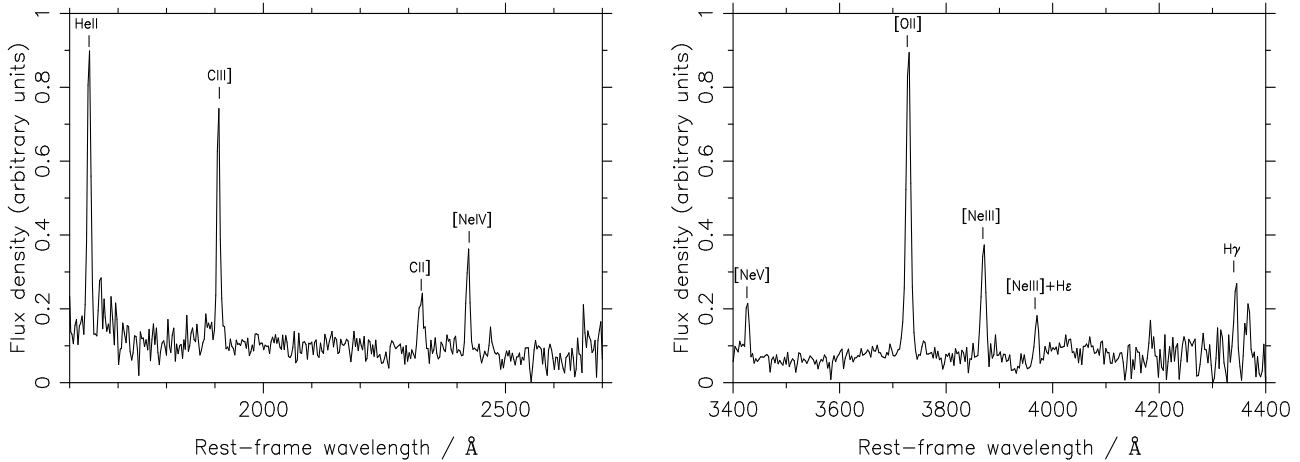

\vspace{2.6 in}
\begin{center}
\includegraphics{c_6C_blue2.ps}
\includegraphics{c_6C_red2.ps}
\end{center}
\caption{Red and Blue arm composite spectra for the seven 6C galaxies.
\label{Fig: 11}}
\end{figure*}
{\bf 6C1217+36} (Fig.~\ref{Fig: 7}) is considerably different from the
rest of the sample. Not only is it the smallest radio source, but it
also apparently does not possess an FRII radio structure (Best {\it et
al}, 1999). The continuum level is slightly greater than the average
for the sample, but the emission lines are very much fainter than
those of any other object. Infrared and optical imaging of this source 
(Inskip {\it et al}, in prep) show a passive giant elliptical galaxy,
its magnitude following the $K-z$ relation (infrared Hubble diagram)
for these sources. The emission lines observed for this source are
more believable in the 2-dimensional spectra (as shown for \oo\, in
frame $c$ of Fig.~\ref{Fig: 7}) than can be displayed here in one
dimension, but are still very faint relative to the continuum
emission.  However, the redshift obtained is in agreement with that
given by Rawlings, Eales and Lacy (2000).  Identifications have
been made for most of the oxygen, carbon and neon emission lines
expected in the spectra.  The two dimensional structure of the \oo\,
line does not appear particularly large in spatial extent, and is in
fact the smallest in the sample. However, this may be due to the poor
S/N in the spectra of this object. The maximum FWHM of this emission line is also the
smallest of the sample, but the very low flux of this line may perhaps
have led to an underestimate of this value when compared with the FWHM
of the remainder of the sample.  Due to the high noise levels and the
fact that this source is atypical of the $z \sim 1$ 6C subsample in
many ways (including its anomalous non-FRII radio structure) this
source has been excluded from any statistical analysis of the larger sample. 

{\bf 6C1256+36} (Figs.~\ref{Fig: 8}--\ref{Fig: 9}) was
observed at five different sky PAs: along the radio axis (40$^\circ$),
and also at angles of 0$^\circ$, 22$^\circ$, 79$^\circ$ and
115$^\circ$.  This was carried out in order to obtain spectra for other
objects in the field. An investigation of a
possible z$\sim$1 cluster will be presented elsewhere. The \oo\,
line of this source is small in spatial extent, and the velocity
profile displays little variation with slit orientation.  The redshift
of this source was found to be 1.128 from our spectra; a previously
published value of 1.07 (Eales {\it et al}, 1997) was an inaccurate estimate
from an at-the-telescope reduction of the Rawlings {\it et al}
spectra.

{\bf 6C1257+36} (Fig.~\ref{Fig: 10}) is the second largest radio
source of the sample, but the spatial extent of the \oo 3727\AA\, line is also
fairly large at about 60kpc. Unfortunately, the \oo\, line of this
source is coincident with a bright sky line, leading to
the low S/N in the two-dimensional emission line. Despite this, the
relative velocity and FWHM vary fairly smoothly with spatial position.
The size of the emission line region is the second largest in the
subsample and, from the results of Best {\it et al} (2000b) for 3CR 
galaxies, this is unexpected for such a large radio source.  However, an
examination of the radio data for this source (Best {\it et al}, 1999)
shows a jet--knot very close to the host galaxy, also associated with a
blue feature in the HST images (Inskip {\it et al}, in prep). The
jet--shocks associated with the production of this hotspot may
induce many small radio source features into the ionization and
kinematic properties of this galaxy. 

\section{Comparison with a matched sample of 3CR galaxies}

\begin{figure*}
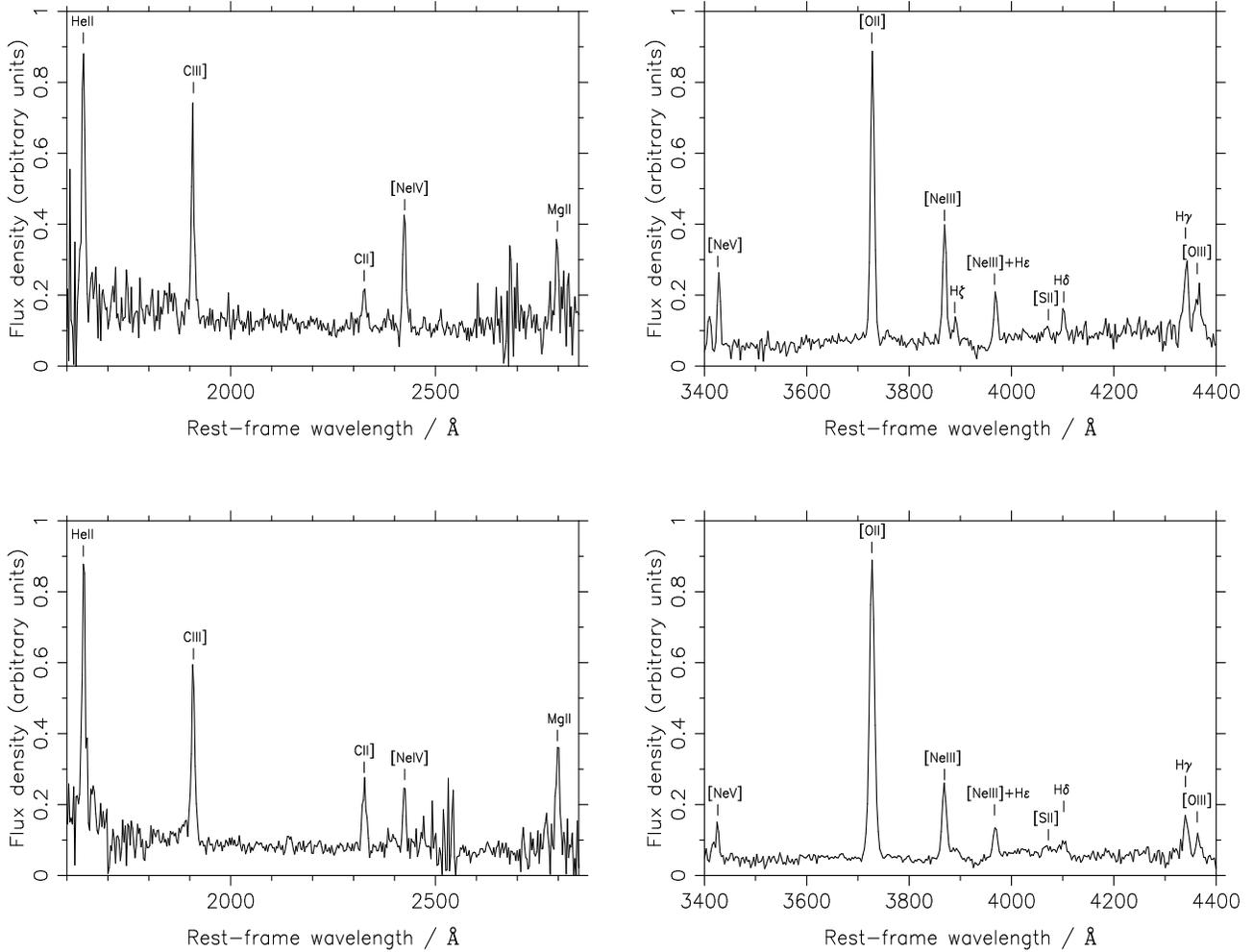

\vspace{5. in}
\begin{center}
\includegraphics{c_BlueAll_pi2.ps}
\includegraphics{c_RedAll_pi2.ps}
\includegraphics{c_BlueAll_si2.ps}
\includegraphics{c_RedAll_si2.ps}
\end{center}
\caption{Composite red and blue arm spectra for both 6C and 3CR
galaxies combined: (a), top -- Large ($>120\,$kpc) radio sources; (b),
bottom -- Small ($<120\,$kpc) radio sources. The
former display many characteristics of photoionized regions, while the
latter spectra are more typical of shock ionization.}
\label{Fig: 12}
\end{figure*}

\subsection{Composite spectra}
Red and blue arm composite spectra have been created for the $z \sim 1$
6C subsample, excluding 6C1217+36 because of its very low
signal--to--noise.  The spectra were combined in the same rest frame,
scaled according to their continuum flux and weighted by the continuum
S/N.  The 6C combined spectra are displayed in
Fig.~\ref{Fig: 11}, and line ratios relative to \oo3727\AA\, are listed
in the fourth data column in Table 3. 
As the flux density scale of the resulting red- and blue-arm composite
spectra is arbitrary, the line ratios in the blue arm relative to
\oo3727\AA\, need calibration. 
This was done by calculating the C\textsc{iii}]1909/\oo3727 ratio, using the mean
C\textsc{iii}]/\oo\, ratio for the entire 6C $z \sim 1$ subsample.

The 3CR subsample of Best {\it et al} (2000a, b) displays considerable
spectral variations between photoionized radio sources (D$_{\rm{rad}}
\gta 120$kpc) and shock ionized sources (D$_{\rm{rad}} \lta 120$kpc).
To investigate these variations in the emission line spectra with
radio size, composite spectra for several different combinations of
sources have been created, in addition to the spectra displayed in
Fig.~\ref{Fig: 11}.  This uses the same weighting system as for the
combination of seven 6C galaxies, scaling the spectra of all sources
by their continuum flux, weighted by the signal--to--noise level of
the continuum.   The line ratios relative to
\oo\, for each of these sets of spectra can be found in Table 3. 
The data sets used consider the 3CR and 6C $z \sim 1$ subsamples
both together and separately; these groups are further subdivided
by radio source size ($<120\,$kpc or $>120\,$kpc).
The C\textsc{iii}]1909/\oo3727 ratio was calculated from the mean
value within each subgroup of galaxies.

Fig.~\ref{Fig: 12} shows the red and blue arm composite spectra for
21 6C and 3CR galaxies, separated into large (usually photoionized)
sources and small (perhaps shock ionized) sources.  There are many
differences between the two sets of spectra. Most emission
lines in the small source composite spectra are fainter relative to
\oo\, than for the large source composite spectra, the Balmer lines
being particularly weak in small sources.  The ratio \crat is also
considerably greater for the larger, photoionized sources.  

\begin{table*}
\caption{Emission line fluxes and associated errors (relative to
\oo3727\AA=100) of the `average' 
spectra for nine different galaxy groupings. Galaxies are grouped by
sample (6C galaxies, 3CR galaxies or all galaxies) and by radio size
(large sources have D$_{\rm{rad}} \geq 120$kpc). Also tabulated is the
scatter in the data, included only when more than three measurements
of a line are available within a particular group of sources. This is
defined as the standard error on the mean value of the scaled line
flux for each group.  The final two rows give the \crat and
\nrat line ratios for the nine different groupings of sources.}
\begin{center} 
\begin{tabular} {ccccccccccc} 
\hline
{Line} && {3CR} & {Small 3CR} & {Large 3CR} & {6C}&{Small 6C}& {Large 6C}&{All}&{All small}&{All large}\\\hline
{C\textsc{iv} 1549} &Flux    & 44  & 42  & --  & 91   & --   & 92   & 45  & 36  & 61 \\
		    &Error   & 6.4 & 6.1 & --  & 13   & --   & 13   & 6.9 & 6.0 & 9.5\\
		    &Scatter & {--}&{--} &{--} & 24\% &{--}  & {--} & 26\%& --  & {--}\vspace*{0.05cm}\\
{He\textsc{ii} 1640}&Flux    & 42  & 36  & 58  & 57   & --   & 70   & 31  & 31  & 43 \\
		    &Error   & 6.2 & 5.2 & 10  & 8.3  & --   & 11   & 4.7 & 4.8 & 7.3\\
		    &Scatter & 24\%&{--} &{--} & 24\% & --   & --   & 17\%& 27\%& 14\%\vspace*{0.05cm}\\
{C\textsc{iii}] 1909}&Flux   & 26  & 23  & 32  & 43   & 42   & 52   & 22  &  22 & 29 \\
		    &Error   & 3.7 & 3.3 & 4.8 & 6.2  & 6.1  & 7.9  & 3.1 & 3.4 & 4.3\\
		    &Scatter & 18\%& 16\%& 30\%& 20\% & --   & --   & 17\%& 16\%& 22\%\vspace*{0.05cm} \\
{C\textsc{ii}] 2326} &Flux   & 8.4 & 8.7 & 7.1 & 14   & 16   & 7.5  & 7.1 & 8.4 &6.7 \\
		    &Error   & 1.4 & 1.7 & 1.1 & 2.3  & 2.4  & 4.1  & 1.1 & 2.5 &1.5\\
		    &Scatter & 22\%&27\% & 24\%& 33\% & --   & --   &18\% &28\% & 20\%\vspace*{0.05cm}\\
{[Ne\textsc{iv}] 2425}&Flux  & 10  & 6.0 & 23  & 20   & 15   & 31   & 9.1 & 6.2 & 20  \\
		    &Error   & 1.7 & 1.3 & 3.4 & 5.2  & 2.3  & 4.9  & 1.4 & 2.4 & 3.2\\
		    &Scatter & 27\%& 33\%& 32\%& 24\% & --   & --   & 19\%&27\% & 24\%\vspace*{0.05cm}\\
{Mg\textsc{ii} 2798} &Flux   & 17  & 17  & 11  & 23   &{--}  & 51   & 14  & 15  & 12.4\\
		    &Error   & 2.7 & 2.6 & 6.3 & 5.0  &{--}  & 11   & 2.1 & 4.0& 3.5\\
		    &Scatter & 35\%&{--} & {--}& {--} &{--}  & {--} & 25\%& 31\%& 38\%\vspace*{0.05cm}\\
{Ne\textsc{v} 3426}  &Flux   & --  & --  &{--} & 14   & 14   & 19   & 11  & 9.5 & 14\\
		    &Error   & --  & --  &{--} & 2.3  & 2.4  & 4.2  & 1.6 & 2.5 & 2.3\\
		    &Scatter &{--} & --  & {--}& 23\% & --   & -- & 21\%& 33\%& --\vspace*{0.05cm} \\
{[O\textsc{ii}] 3727}&Flux   & 100 & 100 & 100 & 100  & 100  & 100  & 100 & 100 & 100 \\
		    &Error   & 10  & 10  & 10  & 10   & 10   & 10   & 10  & 10 & 10\\
		    &Scatter &19\% &22\% &19\% & 29\% & --   & --   & 17\%&20\% & 20\%\vspace*{0.05cm}\\
{[Ne\textsc{iii}] 3869}&Flux & 24  & 21  & 47  & 30   & 28   & 31   & 27  & 22  & 34 \\
		    &Error   & 3.4 & 3.0 & 6.8 & 4.3  & 4.2  & 5.3  & 3.8 & 3.7 & 4.8\\
		    &Scatter & 18\%& 27\%& 19\%& 27\% & --   &--    & 15\%& 23\%& 17\% \vspace*{0.05cm}\\
{H$\zeta$ 3889}	    &Flux    & 3.3 & 3.8 & 8.4 & --   &{--}  & 5.1  & 4.3 & --  & 5.6 \\
		    &Error   & 0.7 & 0.9 & 1.5 & --   &{--}  & 3.8  & 0.6 & --  & 1.3\\
		    &Scatter &17\% & 17\%& 26\%& --   &{--}  & {--} & 16\%& --  & 24\%\vspace*{0.05cm}\\
{H$\epsilon$ + [Ne\textsc{iii}] 3967}&Flux& 10 &10 &14 &7.1&3.9&16  & 12  & 7  & 18  \\
		    &Error   & 1.5 & 1.4 & 2.1 & 1.4 & 2.5   &4.7   & 1.7 &1.8   &3.4 \\
		    &Scatter & 17\%& 21\%&22\% & 18\% & --   & --   &15\% &19\% & 18\%\vspace*{0.05cm}\\
{[S\textsc{ii}]}    &Flux    & 2.4 & 2.4 &3.9  &  --  &{--}  & {--} & 2.8 & 3  &4.5\\
		    &Error   & 1.2 & 1.4 &2.7  &  --  &{--}  & {--} &1.7  &2.0 &2.4\\
		    &Scatter &{--} &{--} &{--} & {--} &{--}  & {--} &{--} &{--} &{--}\vspace*{0.05cm}\\
{H$\delta$ 4102}    &Flux    & 8.1 & 7.0 &12   & --   &{--}  &  --  & 9.1 &6.8 & 11\\
		    &Error   & 1.7 & 1.4 &2.5  & --   &{--}  & --   & 2.3 &2.0 & 3.0\\
		    &Scatter &14\% &{--} &20\% & {--} & {--} & {--} & 16\%& 13\%& 21\%\vspace*{0.05cm}\\
{H$\gamma$ 4340}    &Flux    & 14  & 11  & 31  & 9.0  & 4.9  & 18   & 19  & 13  & 26 \\
		    &Error   & 2.0 & 1.8 &3.4  & 5.0 & 2.7   & 9.7  & 2.8 & 4.1&6.6 \\
		    &Scatter &22\% &37\% &23\% & {--} & {--} & {--} &18\% &30\% & 21\%\vspace*{0.05cm}\\
{[O\textsc{iii}] 4363}&Flux  & 2.4 & 3.6 & 3.9 & --   &{--}  & {--} & 9.4 &4.6 & 11\\
		    &Error   & 1.5 & 2.3 &2.7 & --   &{--}  &{--}  & 2.2  & 2.2  &6.2\\
		    &Scatter & {--}& {--}&{--} &{--}  &{--}  & {--} &{--} &{--} &{--}\vspace*{0.15cm}\\\hline
\multicolumn{2}{c}{\crat}    & 3.10& 2.64& 4.51& 3.07 & 2.63 & 6.93 & 3.10& 2.62& 4.33\\
\multicolumn{2}{c}{Error}    & 0.68& 0.64& 0.97& 0.67 & 0.55 & 3.93 & 0.65& 0.88& 1.16\\
\multicolumn{2}{c}{\nrat}    & {--}&{--} &{--} & 2.14 & 2.00 & 1.63 & 2.45& 2.32& 2.43\\
\multicolumn{2}{c}{Error}    & {--}&{--} &{--} & 0.47 & 0.46 & 0.46 & 0.50& 0.72& 0.53\\\hline
\end{tabular}
\end{center}
\end{table*}

\begin{figure}
\vspace{2.5 in}
\begin{center}
\includegraphics{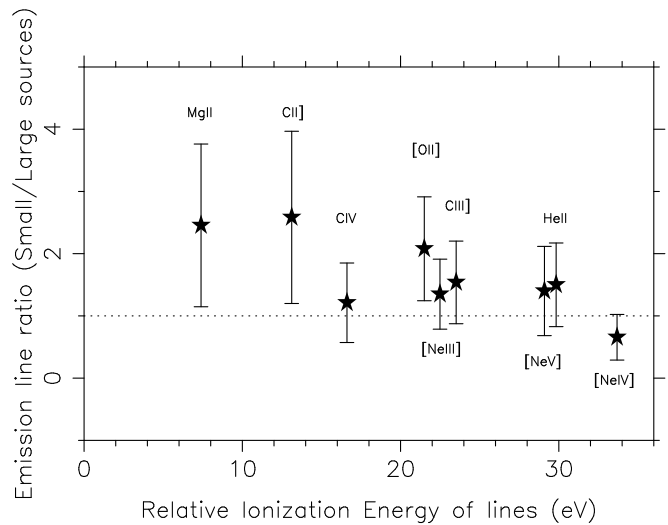}
\end{center}
\caption{Plot of small--to--large source emission line flux ratio
vs. ionization energy of the emission lines, calculated as
the difference in ionization energy between successive ionization
states. The stars represent the results for both 6C and 3CR sources.}
\label{Fig: 13}
\end{figure}

In Fig.~\ref{Fig: 13}, the ratios of the emission line fluxes
of small and large sources are plotted against the ionization energy of each
emission line, calculated as the difference in ionization energy between
successive ionization states (e.g. for C\textsc{iii}, the 
difference between the second and third ionization energies of
carbon).\footnote{Ionization energies are taken from WebElements
(http://www.webele ments.com/) and NIST
(http://physics.nist.gov/cgi-bin/AtData/main\_asd).}   
Ratios are plotted for the combined spectra of both 6C and 3CR
sources. There is a clear
anticorrelation of small/large source flux ratio with ionization state
(95\% significant for the combination of 6C and 3CR galaxies) which
clearly demonstrates the 
change in ionization state with radio size.

\subsection{Ionization of the extended emission line regions}

\begin{figure*}
\vspace{5.2 in}
\begin{center}
\includegraphics{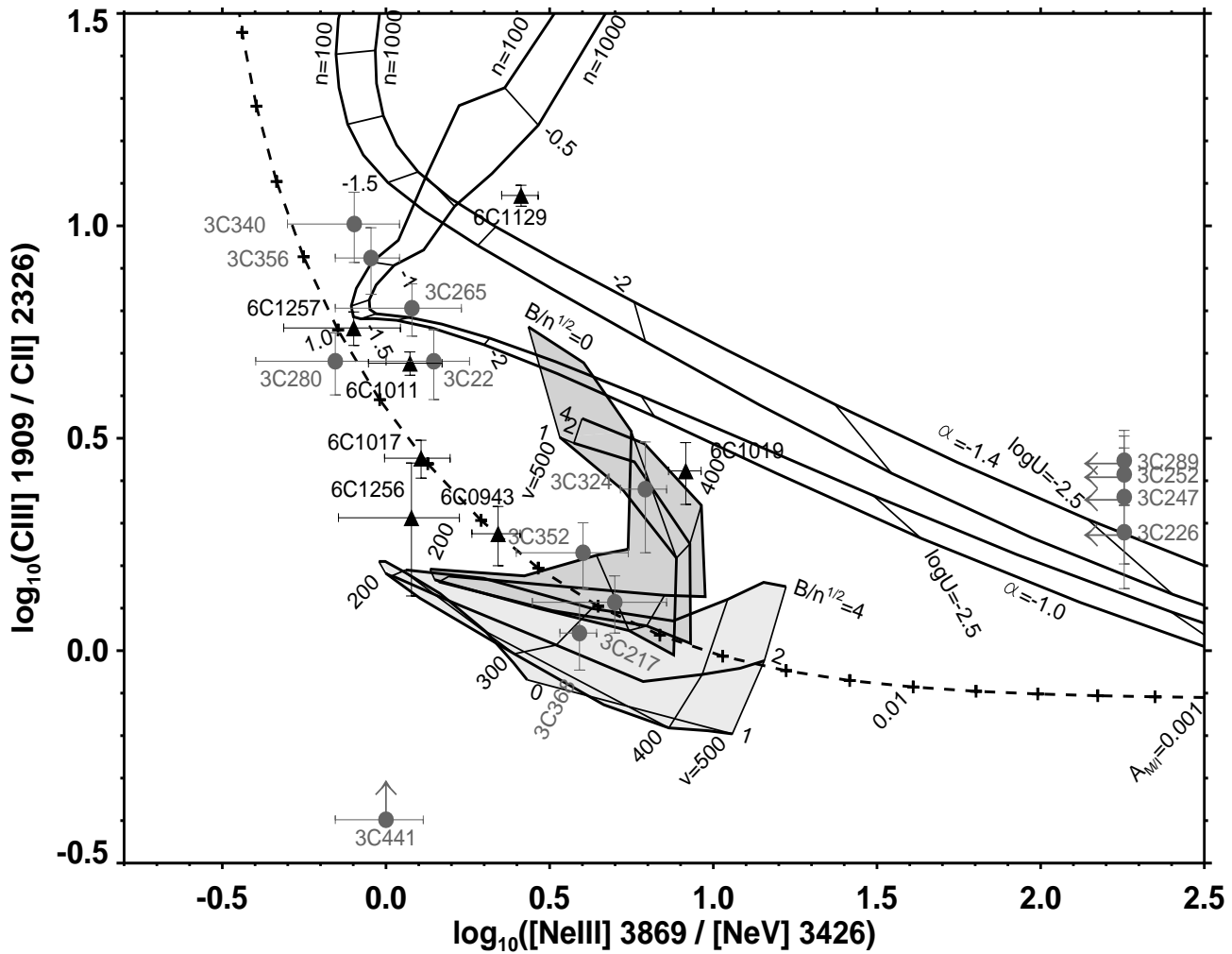}
\end{center}
\caption{A carbon ratio vs. neon ratio emission line diagnostic plot
for 6C (black triangular data points) and 3CR galaxies (grey circular
data points).  For full details of how the theoretical 
lines were determined, see Best {\it et al} 2000b and references
therein.  The results are compared with the theoretical predictions
for shock ionization, simple photoionization, and photoionization
including matter-bounded clouds.  The shock ionization line ratio
predictions are those of the models of Dopita and Sutherland (1996).
Results for both simple shock ionization (faint shading) and also for
models including a precursor ionization region (darker shaded region)
are included on the diagram. 
The precursor ionization region is created by UV photons produced by
the shock which have diffused upstream.  Shock velocities were allowed
to vary between 150 to 500 kms$^{-1}$.  The `magnetic parameter',
B/$\sqrt{n}$ (which controls the effective ionization parameter of the
post-shock gas) was varied from 0 to 4$\mu$G\,cm$^{-1.5}$.  The simple
photoionization model tracks are taken from the theoretical line
ratios of Allen {\it et al} (1998), calculated using the
\textsc{mappingsii} code (Sutherland, Bicknell \& Dopita 1993).  A
power law spectrum illumination ($F_{\nu} 
\propto \nu^{\alpha}$, $\alpha = -1$ or $-1.4$, with a high energy
cut-off at 1.36keV) of a planar slab of material (density $n_e = 100$
or $1000 \rm{cm}^{-3}$) was assumed, with an ionization parameter
$10^{-4}\,\leq\,\rm{U}\,\leq\,1$.  The models correspond to cloud
sizes from 0.003 to 32 parsec, and are ionization bounded.  Evans {\it
et al} (1999) consider the uncertainty in the models of Allen {\it et
al} and Dopita \& Sutherland to be at a level of about 0.3 dex.  Matter
bounded photoionization tracks are also plotted, using the predicted
line ratios of Binette {\it et al} (1996).  Sources plotted at the
edges of the diagram lack data for one of their emission line ratios.  
\label{Fig: 14}}
\end{figure*}

The 6C subsample is well suited for comparison with the 3CR subsample
studied by Best {\it et al.} (2000a, 2000b). Both samples cover a narrow
range of redshift at $z \sim 1$, whilst the 6C sources are
approximately 6 times lower in radio luminosity than the 3CR galaxies.
This enables a direct investigation of the effects of radio power on
the emission properties of radio galaxies to be made. 

Best {\it et al.} made use of a \nrat vs. \crat line diagnostic
diagram to investigate the ionization mechanisms occuring within their
subsample. Complete line ratio information was available for nine of
the fourteen galaxies, and these nine sources clearly fall into either
of two regions on the diagnostic diagram. The two regions are well
matched to the theoretical predictions of shock ionization and
photoionization models, with sources smaller than 115kpc appearing
shock ionized and larger sources lying in the AGN photoionization region 
of the plot.  Fast radiative shocks can also be a strong source of
ionizing photons; this can profoundly affect the ionization state of
the emission line regions, and is included in the shock models used in
comparison with our data.  Throughout this paper, the phrase ``shock
ionization'' generally refers to the models including this additional
photoionizing component.  A plot of the emission line ratio
(C\textsc{iii}]1909/C\textsc{ii}]2326) against radio size showed a
clear correlation between the two parameters, significant at the
98.5\% level. The interpretation of these results was that the
ionization state of the emission line gas varies strongly with radio
size, and that the dominant ionization mechanism is photoionization
for large radio sources ($D_{\rm{rad}} > 120 \rm{kpc}$) and shock
ionization for smaller sources ($D_{\rm{rad}} < 120
\rm{kpc}$). The equivalent widths of the \oo 3727\AA\, line are greater in
size for small sources, and their continua also appear somewhat
brighter on average.  For the nine sources with complete line ratio
data, a comparison of the average integrated flux of the \oo\, line for
photoionized and shock ionized sources suggested that the power of the
\oo\, line is boosted in smaller, shocked sources. 

Fig.~\ref{Fig: 14} shows the line ratio diagnostic diagram of Best
{\it et al}, including the theoretical predictions and 3CR data. To
this we have added the results obtained from our 6C subsample, as
tabulated in Table 4. Whilst the 3CR galaxies marked on the line
diagnostic diagram fall neatly into the two well defined regions of
either shock ionization or photoionization, the same cannot be said
for the 6C galaxies. Based on radio source size, 6C0943+39, 6C1019+39,
6C1011+36, 6C1256+36 and 6C1257+36 lie within the `expected' regions
on the diagnostic plot. The position of 6C1129+37 (a radio source of
intermediate size) is unusual, located a fair distance from the major
group of photoionized sources. However, imaging observations of this
source suggest that it may be interacting with a nearby companion
galaxy (Inskip {\it et al., in prep}).  The small radio source
6C1017+37 lies between the predictions of photoionization and shock
ionization models.  Although the size vs. emission line ionization
mechanism relationship for the 6C galaxies is not as clear as for the
3CR galaxies, there is a similar trend.   The photoionization models
of Binette {\it et al} (1996) also appear to fit the data fairly well;
further diagnostic plots utilising other line ratios are required in
order to determine the most appropriate models.

\begin{table*}
\caption{Ionization and kinematic properties of the emission line
regions of the 6C radio galaxies. Column 1 gives the name of the radio
source, and column 2 its redshift. The projected linear size of the
radio source is tabulated in column 3. Columns 4 to 7 give the
C\textsc{iii}]1909/C\textsc{ii}2326] and
[Ne\textsc{iii}]3869/[NeV]3426 emission line ratios and their errors.
Column 8 contains the integrated \oo 3727\AA\, flux density, and
column 9 its rest frame equivalent width. The physical size of the
emission line region as determined from the extracted two dimensional
image of the \oo 3727\AA\, emission line is given in Column
10. Columns 11 to 13 also contain information on the
\oo 3727\AA\, emission line: its maximum FWHM, the observed range
in relative velocities along the slit, and the number of
discrete velocity components. }
\begin{center} 
\begin{tabular} {lcccccccccccc} 
\hline
{Source}&{z}&{Radio}&\multicolumn{2}{c}{C\textsc{iii}]/C\textsc{ii}]}&\multicolumn{2}{c}{[Ne\textsc{iii}]/[NeV]}&{\oo
flux}&{Eq.}&{Emis.}&{Max.}&{Vel.}&{No. comps.}\\
{ }   &{ } & {size} &
{ratio}&{error}&{ratio}&{error}&{($\times10^{-19}$)}&{width}&{size}&{FWHM}&{Range}
& {(N$_v$)}\\
{ }   &{ } & {[kpc]} &{ } &{ } &{ } &{ } & {[$\rm{W\,m}^{-2}$]}&{[\AA]}&{[kpc]}&{[km/s]}&{[km/s]}&{[kpc]}\\
{(1)}      & {(2)} & {(3)}& {(4)}  & {(5)} & {(6)}  & {(7)} & {(8)} & {(9)}& {(10)}& {(11)}& {(12)}&{(13)}\\\hline
{6C0943+39}&{1.036}& {92} & {1.885}& {0.30}& {2.200}& {0.37}& {5.01}& {211}& {68.4}& {860} & {305} & {2-3?}\\
{6C1011+36}&{1.042}& {444}& {4.749}& {0.30}& {1.184}& {0.30}& {1.24}& {41} & {18.6}& {400} & {80}  & {1}\\
{6C1017+37}&{1.053}& {65} & {2.838}& {0.29}& {1.280}& {0.29}& {6.86}& {195}& {43.7}& {880} & {245} & {2}\\
{6C1019+39}&{0.922}& {67} & {2.649}& {0.44}& {8.238}& {0.94}& {2.43}& {58} & {42.3}& {1300}& {475} & {3$^1$}\\
{6C1129+37}&{1.060}& {141}& {11.79}& {0.67}& {2.586}& {0.33}& {5.14}& {226}& {56.2}& {600} & {300} & {2?}\\
{6C1217+36}&{1.088}& {38} & {1.208}& {1.02}& {1.000}& {1.22}& {0.51}& {59} & {12.6}& {265} & {20}  & {1}\\
{6C1256+36}&{1.128}& {155}& {2.055}& {0.71}& {1.196}& {0.48}& {2.09}& {147}& {31.6}& {660} & {40}  & {1}\\
{6C1257+36}&{1.004}& {336}& {5.750}& {0.52}& {0.797}& {0.31}& {2.24}& {83} & {61.7}& {775} & {60}  & {1}\\
\hline
Notes: &{[1]}&\multicolumn{11}{l}{This source also contains a discrete
high velocity component.}\\
\end{tabular}
\end{center}
\end{table*}

\begin{table*}
\caption{Correlation Table. The values tabulated give the significance
level of any correlation found using a spearman rank test. A `*'
denotes correlations at a significance level of 75\% or less, and
implies an uncorrelated pair of data sets. Positive and negative signs indicate whether a parameter
set is correlated or anti-correlated respectively.  Also included are
the probabilities of the results of the $\chi^2$ analysis of the
distribution of the N$_v$
parameter as a function of radio size occurring by chance.}
\begin{center} 
\begin{tabular} {lr@{.}lccr@{.}lccr@{.}lcc} 
\hline
{Parameters} 		 & \multicolumn{4}{c}{3CR data} & \multicolumn{4}{c}{6C data} & \multicolumn{4}{c}{All data} \\
 &\multicolumn{4}{c}{} &\multicolumn{4}{c}{excluding 6C1217+36}&\multicolumn{4}{c}{excluding 6C1217+36} \\\hline
                 & \multicolumn{2}{c}{$r_s$} & n & probability&  \multicolumn{2}{c}{$r_s$} & n & probability&  \multicolumn{2}{c}{$r_s$} & n & probability \\
{Carbon Ratio vs. Radio Size}		 & 0&681 & 13 & +99.5\% & 0&357 & 7 & +78\%  & 0&523 & 20 & +99.1\%\\
{Maximum FWHM vs. Radio Size}		 &-0&650 & 14 & -99.4\% &-0&821 & 7 & -98.8\%&-0&667 & 21 & -99.95\%\\
{Carbon Ratio vs. Maximum FWHM}	 &-0&667 & 13 & -99.4\% &-0&464 & 7 & -85\%  &-0&537 & 20 & -99.3\%  \\
{EELR Velocity range vs. Radio Size}	 &-0&282 & 14 & -84\%   &-0&631 & 7 & -93.6\%&-0&330 & 21 & -92.8\%  \\
{EELR Velocity range vs. Maximum FWHM}	 & 0&408 & 14 & +92.6\% & 0&487 & 7 & +87\%  & 0&480 & 21 & +98.6\%  \\
{Scaled [OII] Luminosity vs. Radio Size} &-0&437 & 14 & -94.1\% &-0&757 & 7 & -97.6\%&-0&528 & 21 & -99.3\%\\
{EELR Size vs. Radio Size }		 &-0&286 & 14 & -84\%   &-0&286 & 7 & *      &-0&200 & 21 & -81\%  \\
{          [+25\% size]}                 &-0&286 & 14 & -84\%   &-0&286 & 7 & *      &-0&268 & 21 & -88\%  \\
{Scaled EELR Size vs. EELR Velocity range} & 0&718 & 14 & +99.8\%& 0&288 & 7 & *      & 0&578 & 21 & +99.7\%  \\\\
{N$_v$ distribution probability}	 & \multicolumn{4}{c}{$<1$\%}&\multicolumn{4}{c}{$<1$\%}   &\multicolumn{4}{c}{$<0.1$\%}\\
\hline
\end{tabular}
\end{center} 
\end{table*}

\begin{figure}
\vspace{2.2 in}
\begin{center}
\includegraphics{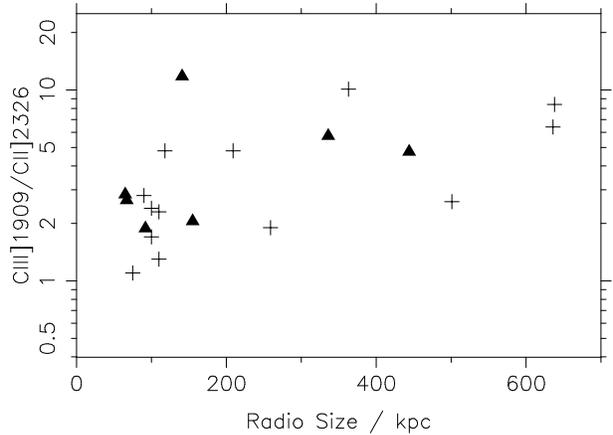}
\end{center}
\caption{The correlation between the \crat emission line ratio and the
projected linear size of the radio source. 6C galaxies are marked as
filled triangles, and the 3CR galaxies studied by Best {\it et al} are
marked as crosses. 
\label{Fig: 15}}
\end{figure}

In Fig.~\ref{Fig: 15} the carbon line ratio is plotted against radio
size for the galaxies in both subsamples. These are very strongly
correlated at the 99\% significance level using a Spearman Rank
correlation test (see Table 5) for all galaxies in both samples, at
the 99\% level for 3CR galaxies alone and less strongly at the $>$75\%
level for the 6C sources alone.  The trend in ionization state is
present for both samples, although somewhat weaker for the smaller
subsample of 6C sources.  This same trend is observed at higher
redshifts: de Breuck {\it et al} (2000) find that the \crat line ratio
is correlated with radio size for their sample of higher redshift
galaxies.  The observed increase in \crat with radio source size is
also mirrored by the results of Jarvis {\it et al} (2001) at $z > 2$,
for which composite spectra have been made for sources $>
70\,\rm{kpc}$ and $< 70\,\rm{kpc}$ in radio size. These spectra
clearly show a marked increase in the \crat line ratio in the larger
sources.   
Stern {\it et al} (1999) provide a composite spectrum for 13 high
redshift ($1.25 < z < 3.6$) radio galaxies, selected from the
MIT-Green Bank (MG) survey. These sources have flux densities 
intermediate between the 6C and 3CR sources at those redshifts, and
radio sizes $< 90$kpc.  The \crat line ratio for these galaxies is 1.2, which
matches the values obtained for the smaller 6C and 3CR sources, and
fits the predictions of the shock ionization models plotted on
Fig.~\ref{Fig: 14}. 
It is also interesting to compare these line ratios with those of radio
quiet quasars, which do not possess radio jets and so their emission line
spectra should be dominated by photoionisation from the AGN. Composite
spectra for QSOs show \crat line ratios of 5.3 (Boyle 1990) and 5.0
(Francis {\it et al} 1991), indeed as expected for AGN photoionization and
comparable to the values obtained for large 6C and 3CR radio sources.

\subsection{Morphological and kinematical properties of the
emission line gas}

The 3CR galaxies studied by Best {\it et al} displayed some
distinctive kinematic characteristics:
\begin{enumerate}
\item{}The velocity width of the 3CR galaxies was shown to be strongly
anti-correlated with radio size at greater than the 99\% significance
level.
\item{}Small radio sources, defined as having a projected linear size
less than 150\,kpc, generally have more distorted velocity profiles
than their larger counterparts, which have velocity profiles
consistent with rotation. The number of separate velocity components was
quantified by the parameter $N_{v}$, defined as the number
of single gradient velocity components required to fit the velocity
profile along the slit direction.   This parameter displayed a clear
variation with radio size, with all but one of small radio sources
having values $N_{v}\,>\,1$ (distorted velocity profiles), and six out
of seven of the larger source having $N_{v}\,=\,1$ (quiescent
rotation-like profiles). $\chi^{2}$ testing showed that the
probability of this occurring by chance was less than 1\%. 
\item{}The emission line regions of small radio sources are greater
than those of larger sources.
\end{enumerate}

\begin{figure}
\vspace{5. in}
\begin{center}
\includegraphics{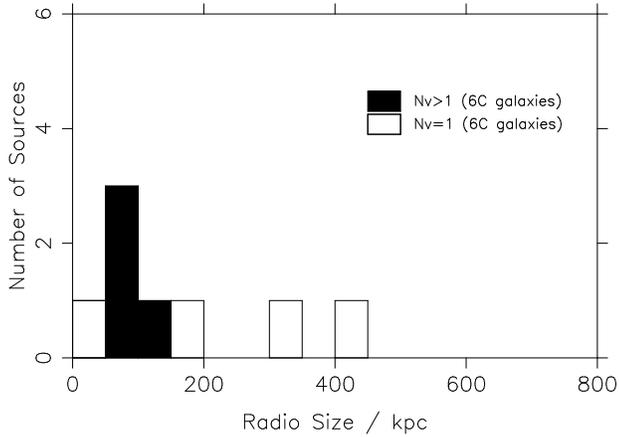}
\includegraphics{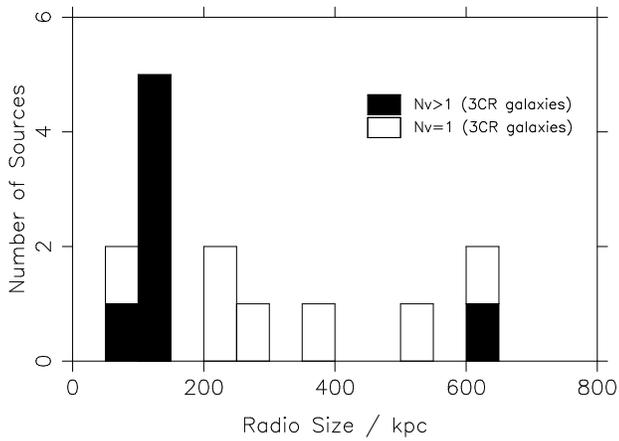}
\end{center}
\caption{(a) The distribution of radio
source sizes of 6C galaxies with smooth velocity profiles (white) and
irregular velocity profiles (black). (b) The same results for the 3CR
galaxies studied by Best {\it et al} (2000b).  The slight
difference with respect to the diagram presented by Best {\it et al}
arises only from the different cosmology adopted in this paper.
\label{Fig: 16}}
\end{figure}
\begin{figure}
\vspace{2.2 in}
\begin{center}
\includegraphics{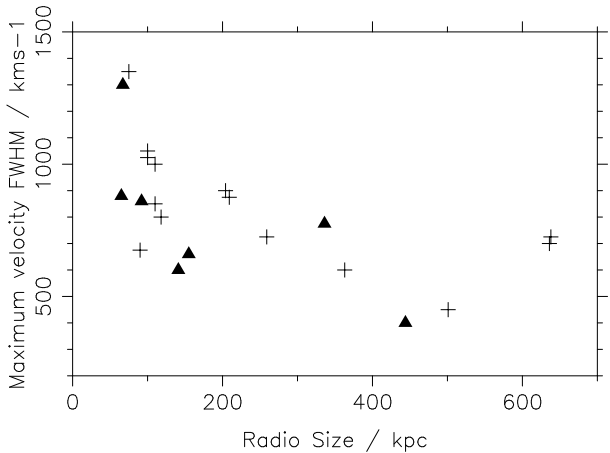}
\end{center}
\caption{The inverse correlation between the maximum FWHM of the
\oo 3727\AA\, emission line and the projected linear size of the radio
source. 6C galaxies are marked as
filled triangles, and the 3CR galaxies studied by Best {\it et al} are marked as
crosses. 
\label{Fig: 17}}
\end{figure}
\begin{figure}
\vspace{2.2 in}
\begin{center}
\includegraphics{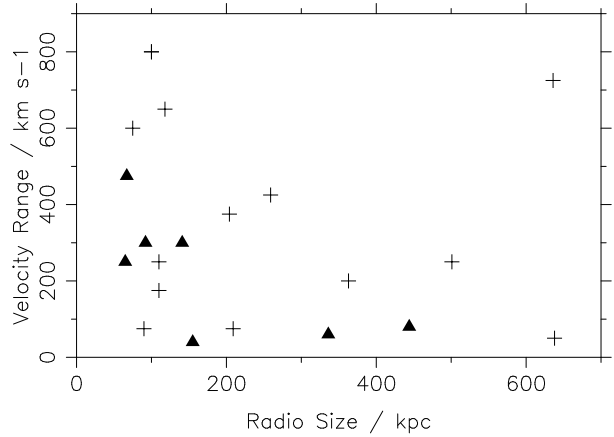}
\end{center}
\caption{The variation of the velocity range observed for the emission
line region with the size of the radio source. 6C galaxies are marked as
filled triangles, and the 3CR galaxies studied by Best {\it et al} are marked as
crosses. 
\label{Fig: 18}}
\end{figure}
\begin{figure}
\vspace{5. in}
\begin{center}
\includegraphics{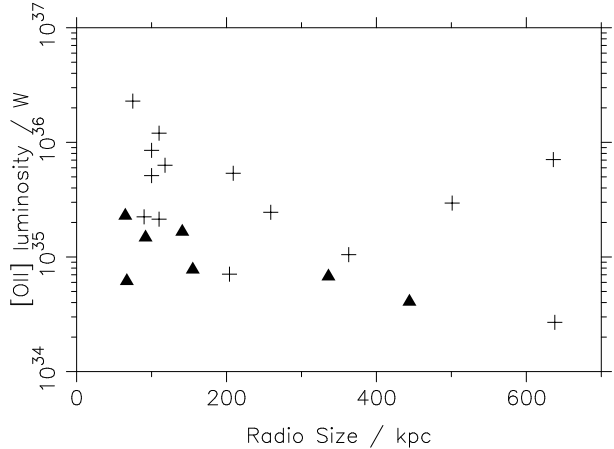}
\includegraphics{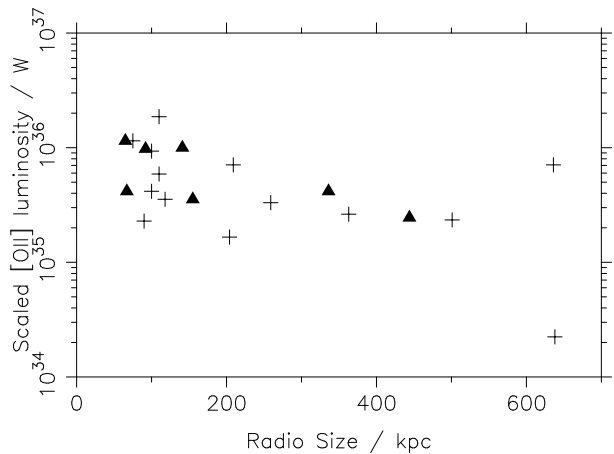}
\end{center}
\caption{The variation of the luminosity of the \oo 3727\AA\,
emission line with the size of the radio source. (a -- top) Unscaled
line luminosities. (b -- bottom) Line luminosities scaled by radio
power at 151MHz as a fraction of the average 3CR radio power. 6C galaxies are marked as
filled triangles, and the 3CR galaxies studied by Best {\it et al} are marked as
crosses. 
\label{Fig: 19}}
\end{figure}

These results can be explained by the effects of shocks passing
through the emission line gas in the smaller sources, and show a
direct connection between the ionization and kinematical properties of
the emission line regions of these galaxies.  The equivalent kinematic
properties of the eight 6C radio galaxies are tabulated in Table 4,
namely the integrated \oo\, emission line intensity and equivalent
width, the projected size of the emission line region, the maximum
FWHM, the range in relative velocities and the number of discrete
velocity components observed in the \oo\, line.  We have investigated
how well the relations found for the 3CR subsample also hold true for
the less powerful sample of eight 6C radio galaxies.  The significance
of any correlations found is tabulated in Table 5.

Figure \ref{Fig: 16}a displays the dependence of N$_v$ on radio size
for the 6C galaxies currently investigated, and Figure \ref{Fig: 16}b the
3CR galaxies of the matched sample previously studied by Best {\it et
al.} The 6C galaxies follow the same trend as 
those in the 3CR subsample: a $\chi^{2}$ analysis of the data shows
that the probability that this distribution occurs by chance is less 
than 1\% for the 6C sample, and less than 0.1\% for both samples
combined (Table 5).

The 6C galaxies are also indistinguishable from the 3CR galaxies in
their variation of maximum velocity FWHM with radio size
(Fig.~\ref{Fig: 17}). An anticorrelation of FWHM with  
ionization state, as quantified by the \crat ratio, is also found.
Another parameter which behaves similarly with radio size for both
3CR and 6C sources is the measured range in velocities observed in the
\oo\, line (Fig.~\ref{Fig: 18}).  This is defined as the difference
between the most positive and negative velocity components of the \oo\,
3727\AA\, line as shown in figures 1-10(e), excluding all data points with
uncertainties greater than 100\,kms$^{-1}$.  On the whole, the
velocity range data is weakly anticorrelated with radio size.  3C265
(the isolated point at the top right corner of Fig.~\ref{Fig: 18}
possesses a very bizarre optical morphology (Best, Longair,
R\"{o}ttgering 1997).  The observed morphology and high velocity gas
may be due to a cooling flow or the aftermath of a galaxy merger,
although the exact nature of this source remains a mystery.  With the
exception of this unusual source, there is a clear tendency for
larger sources to have a narrower range of velocities than
smaller sources.  However, the first difference between
the two samples appears in this diagram.  The lower radio power 6C
galaxies appear to be restricted to a narrower, less extreme range of
velocities than the more powerful 3CR sources, at all radio sizes.
Kolmogorov-Smirnov testing of these trends shows that the differences
observed (between large and small sources and between the two samples)
are statistically significant at only the 74\% and 51\% levels
respectively.  A larger sample would be required to confirm (or
disprove) the statistical significance of this result. 
Velocity range is also positively correlated with the FWHM of the
emission line gas at the 90\% significance level for the combination
of both samples.

Fig.~\ref{Fig: 19} displays the variation of the \oo\, line luminosity
($L_{\rm{[OII]}}$) with radio size for both samples.  This is more
appropriate than the \oo\, equivalent width (used by Best {\it et al})
for the 6C galaxies, for several reasons.  Due to the extremely low
flux levels of the continua of the 6C galaxies, the measurements of
the \oo\, line equivalent widths are subject to large uncertainties.
Also, the 3CR data show considerable boosting of the line flux of the
smaller, shock ionized sources.  This effect seems more important in the
6C sources, especially 6C1217+36 and 6C1019+39, which have
particularly low equivalent widths.  More importantly, if we wish to
compare the 3CR and 6C sources we need to take into account the fact
that there is roughly a factor of 5 difference in radio power between
3CR and 6C galaxies.  

Rawlings and Saunders (1991) showed that radio jet kinetic power is
strongly correlated with narrow line luminosity in radio sources. Baum
\& Heckman (1989) find a strong correlation between radio and narrow
line luminosities. Jarvis {\it et al} (2001) and Willott {\it et al}
(1999) also provide evidence for a positive correlation between these
luminosities; their results suggest that this relation is close to
being a proportionality, although the exact slope is still unclear.  
From this correlation, the line emission of the 6C galaxies is
expected to be significantly weaker than that of more powerful 3CR
galaxies. However, the relation of continuum emission to radio power
is not so well known.  Without this, we cannot gauge the reliability
of comparing equivalent widths of two data sets at different radio
powers, and so we compare the luminosities instead.  In order to be
able to directly compare the two samples, Fig.~\ref{Fig: 19}b shows
the line luminosity of each source scaled by the ratio of the host
galaxy's radio flux density at 151MHz to that of the average for the
3CR subsample.  Assuming that radio and narrow line luminosities are
proportional, this removes to first approximation the decrease in
$L_{\rm{[OII]}}$ in the 6C sources due to the difference in radio
power between the two subsamples.  The resulting \oo\, line
luminosities of galaxies from both samples occupy a narrow range of
values, and are anticorrelated with radio size.  This anticorrelation
is significant at the 97\% level for the 6C subsample alone, 94\% for
the 3CR subsample alone, but at greater than the 99\% significance
level when the samples are combined.  

Interpretation of the $P$--$D$ relation (Scheuer 1974; Baldwin 1982)
for radio galaxies predicts that for an assumed constant AGN output,
the radio luminosity of a source will decrease monotonically over its
lifetime (Kaiser {\it et al} 1997).  If two sources are observed to have
identical radio luminosities, the larger source of the two will host
an intrinsically more powerful AGN, and thus produce more luminous
\oo\, emission due to the strong correlation between radio power and
emission line luminosity.  By scaling our data to the same radio
luminosity, we can expect the prediction of greater emission line
luminosity for larger, more  powerful sources to reduce the strength
of any intrinsic correlation between $L_{\rm{\oo}}$ and radio source
size/age.  The observed anticorrelation therefore provides a lower
limit to the strength of the actual anticorrelation between
$L_{\rm{\oo}}$ and radio source age. 

An examination of the line intensity vs. spatial offset plots for the
6C galaxies, in contrast to the equivalent plots for 3CR sources by
Best {\it et al.} (2000a), shows that fits to the \oo\,  emission
lines of 6C galaxies generally result in a more steeply edged line
intensity profile. This is a result of the much lower S/N in the 6C
spectroscopic data, despite the longer integration times of the
observations.  We are unable to fit Gaussians to the extracted
two-dimensional \oo 3727\AA\, emission line out to the same spatial
extent as would have been possible with higher signal-to-noise data,
and as this is the method by which we determined the spatial size of
the emission line regions, comparing different EELR sizes within each
of the two data sets is not straightforward.  

\begin{figure}
\vspace{2.2 in}
\begin{center}
\includegraphics{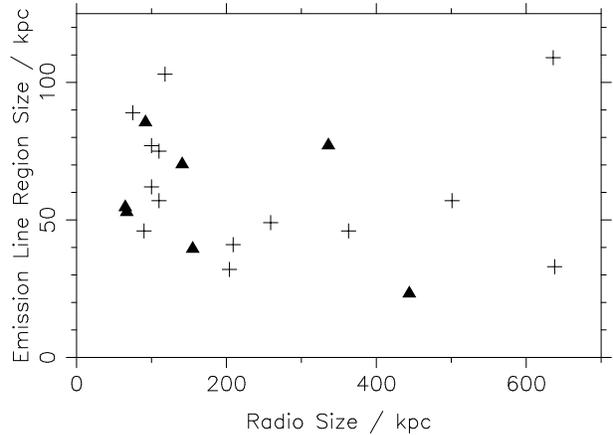}
\end{center}
\caption{The variation of the linear extent of the \oo 3727\AA\,
emission line region with the size of the radio source, once
corrections for the lower S/N ratio of the 6C data have been included.
6C galaxies are marked as filled triangles, and the 3CR galaxies
studied by Best {\it et al} are marked as crosses. 
\label{Fig: 20}}
\end{figure}

To overcome this problem, we added noise to the two dimensional region
about the \oo 3727\AA\, emission line for 3CR sources, reducing the
S/N to a level equivalent to that observed in the 6C sources. By
determining the spatial extent of the region in which fits to the
\oo\, line could be made in this noisier version of the spectra, we
estimated the amount by which the size of the emission line regions in
6C sources may have been underestimated.  Typically, the addition of
noise to the two-dimensional 3CR spectra resulted in a reduction in
the deduced size of the emission line region of between 15--30\%.
Fig.~\ref{Fig: 20} plots emission line region size vs. radio size,
after the former has been scaled up by 25\% for the 6C sources.  It
can be seen that after accounting for this reduced S/N of the 6C
sources, the relation between emission line region size and radio size
for these galaxies is in agreement with that found by Best {\it et al}
for the 3CR subsample.  We also see that the size of the emission line
region is strongly correlated with the range of velocities observed in
the emission line gas for the combination of the two samples, although
only weakly for the 6C sources alone.  The lack of an anticorrelation
in the 6C data is due in part to the position of 6C1257+36 on
Fig.~\ref{Fig: 20}, which may perhaps be currently undergoing a
jet-cloud interaction, and this may explain its larger than expected
EELR size. 

One other point to note is that the two dimensional extracted spectra
of the \oo 3727\AA\, lines which have the most distorted profiles,
seemingly indicative of shocks, occur only in the galaxies whose
ionization states place them in the shock region of the line diagnostic
diagram.  We also see that the smoothly profiled sources occupy the
photoionization region of the plot.  The small radio source 6C1017+37
has a large emission line region, typical for its radio size, but has
a very smooth profile more like that found in the larger, photoionized
radio galaxies.  This source lies within the photoionization region of
the emission line diagnostic plot. Although radio source size is a 
fairly reliable indicator of ionization state (and perhaps ionization
mechanism), the appearance of the \oo 3727\AA\, line also appears to
correlate with the best fitting model of the ionization mechanism
causing the excitation of the emission line regions.  This result also
strongly suggests that the ionization and kinematic properties of the
EELRs are fundamentally linked.

\section{Discussion}

A consistent picture emerges from the analysis of the 3CR and 6C
observations discussed in Section 4. In summary, 

\begin{itemize}

\item[(a)]the small radio sources ($D_{\rm{rad}} < 120\,\rm{kpc}$) have
extensive emission line regions with distorted velocity profiles,
indicating multiple components present in the gas, and have line ratios
consistent with the predictions of models for ionization by fast
radiative shocks.  Typically,
these sources possess emission line regions similar in size to the radio
source, up to 100\,kpc.  The greater \oo\, luminosities of these sources
compared with the large radio sources is in part due to the increased size
of their emission line regions.

\item[(b)]The larger radio sources ($D_{\rm{rad}} > 120\,\rm{kpc}$) have
compact emission line regions with smooth velocity profiles, and their
emission line ratios are well matched by the predictions of the
Allen {\it et al} (1998) models for photoionization by an obscured
AGN.  For these sources 
photoionized gas is observed out to similar extents from the AGN ($\sim
25\rm{kpc}$) in both the 6C and 3CR samples, once the differing
signal-to-noise ratios of the observations have been taken into account.
The photoionization models which best fit the emission line ratios of the
less powerful 6C radio sources have on average lower ionization parameters
(defined as the ratio of the number density of ionizing photons to the
number density of the gas) than those found in the large 3CR sources.

\end{itemize}

Although it is evident that the two different types of excitation each
dominate in a subset of the sources, Fig.~\ref{Fig: 14} shows that the
sources lie along a continuous sequence in the line-ratio diagram and it
is almost certain that both processes play a role in the majority of the
sources, with the balance between the two changing with radio size.

\subsection{The ionization and structure of the photoionized extended
emission line regions} 

For large, photoionized radio sources, the total emission line flux scales
with radio power for the two samples (Fig. 19) and the physical extent of
the emission line regions are comparable ($\sim 25$\,kpc) at different
radio powers. These results can be accounted for if the clouds making up
the EELR are ionization bounded with a small covering factor; their
minimum sizes can then be estimated using the expression for the H$\beta$
line luminosity given by Osterbrock (1989):

\begin{equation*}
\begin{split} 
L_{\rm{H}\beta} 
&= h\nu_{\rm{H}\beta} \frac{\alpha^{\rm eff}_{\rm{H}\beta}(H^0,T)}{\alpha_B(H^0,T)}\int_{\nu_0}^\infty\frac{L_\nu}{h\nu}\,{\rm d}\nu\\[5pt]
&= n_e\,n_p\alpha^{\rm eff}_{\rm{H}\beta}h\nu_{\rm{H}\beta}V\epsilon,
\end{split}
\end{equation*}
where $\alpha^{\rm eff}_{\rm{H}\beta}(H^0,T)$ and $\alpha_B(H^0,T)$
are the recombination coefficients for the $H\beta$ line and the total
recombination rate, $V$ is the total volume and $\epsilon$ is the
filling factor ($\epsilon = 1$ for an individual cloud).

If the flux of ionizing photons reaching the cloud gives an ionization
parameter $U$, $Ucn_p = \int_{\nu_0}^\infty (L_\nu/{h\nu})\,{\rm
d}\nu$.   Then, the depth to which the cloud would be fully ionized
would be  $d = Uc / n_e\alpha_B(H^0,T)$.  Taking $\log_{10}U = -1$
(from the analysis of the photoionized galaxies as plotted on
Fig.~\ref{Fig: 14}) and $T \sim 10^4$ K, the minimum size of an
ionization bounded cloud would be $d \sim 37.5\,(10^8{\rm
m}^{-3}/n_e)$ pc.  These figures are consistent with the standard
picture in which the photoionized clouds occupy only a small fraction
of the observed emission line region.  

\subsection{Radiative shocks in small radio sources}

The emission line ratios of sources with radio sizes $D_{\rm{rad}} <
120\,\rm{kpc}$ can be explained by the predictions of the shock ionization
models of Dopita \& Sutherland (1996), in which certain emission lines are
boosted by shock excitation. The changing line ratios as compared with
photoionization models which successfully account for the spectra of the
large sources, are evidence that shock boosting is important in the small
radio sources.  For example, the C\textsc{iii}]1909/C\textsc{ii}]2326 line
ratio changes systematically with radio size. A similar trend is seen in
the spectra of radio sources at much larger redshifts (de Breuck {\it et
al.} 2000; Jarvis {\it et al.} 2001).  Another characteristic of the
spectra of the smaller radio sources, observed in both $z \sim 1$
subsamples, is a reduction in Balmer line flux. Morse, Raymond and Wilson
(1996) have noted that the high electron temperature produced by fast
ionizing shocks reduces the efficiency of Balmer emission relative to
emission by UV semi--forbidden lines and resonance lines, indicating that
this result too is consistent with shock ionization playing an important
role. Let us therefore study the physics and energetics of the shock
excitation models in a little more detail, to consider whether or not the
shock energetics are indeed sufficient to explain the observed properties
of the emission line regions.

In considering the effects of the shocks, we must first consider the
properties of both the shock and the gas into which it is expanding.  We
consider (cf. review by McCarthy 1993) an elliptically-shaped shock wave
associated with the cocoon of the expanding radio source, propagating at a
velocity of 0.01\,--\,0.03$c$ through the ambient diffuse hot ($\sim 10^7$
to $10^8$K) gas surrounding a massive radio galaxy.  Cool ($\sim 10^4$K)
dense clouds are embedded in the hot gas with very small filling factor
($\sim 10^{-6}$); these are in pressure equilibrium with the hot phase,
implying a density ratio of $n_{\rm cloud}\,/\,n_{\rm IGM} \sim 10^3$ to
$10^4$ . It is these cooler clouds that produce the line emission.

The passage of highly supersonic shocks through the cool clouds can have a
profound effect on their temperature, density and ionization state.  The
shock models of Dopita \& Sutherland include the presence of magnetic
fields to prevent excessive compression behind the shock, and predict that
the shocked gas is compressed as $n_f/n = (8\pi\times10^{-7}\mu
m_H)^{1/2}V_{\rm s}/(B/n^{1/2})$, which for magnetic flux densities of $B
\sim 1-5 \rm{nT}$, determined from estimates of the equipartition field
strengths in the radio source cocoons, corresponds to a compression factor
of $\sim 10^2$. Because of the compression of the gas, the ionization
parameter for photoionization decreases. The effects of cloud compression
or expansion alone cannot, however, account for the observed differences
between the ionization states of large and small sources; for example, the
reduction of $\log_{10}U$ from -1 to -3 that would be required for
photoionization to account for the observed change in the \crat line ratio
between large and small radio sources, would not be compatible with the
changes in the observed \nrat line ratio (the implied change in
$\log_{10}U$ from this line ratio is $\sim 0.7$ to 1; see Fig.~\ref{Fig: 14}). 
Furthermore, excessive compression of the gas clouds could lead to
collisional de--excitation.  Since the critical density at which \oo
3727\AA\, is collisionally de--excited is $\sim 10^{10}\,{\rm m}^{-3}$
(Morse, Raymond \& Wilson 1996), the compressed gas cannot exceed this
density. 

The shock models of Dopita \& Sutherland (1996), which assume that all
the mechanical energy of the shock is radiated,
provide a scaling relation
for the total radiative luminosity per unit area of a shock,
\begin{equation*}
L_{\rm T} = 2.28 \times 10^{-6}\,(V_{\rm s}/100\, {\rm km}\,{\rm s}^{-1})^{3.0}\,(n/10^6{\rm m}^{-3})\,{\rm W}\,{\rm m}^{-2},
\end{equation*}
where $V_{\rm s}$ is the shock velocity and $n$ the ambient density of the
pre--shock gas. The shocks propagate at a much lower speed through the
clouds than through the inter-cloud hot gas, due to their greater
density. Assuming pressure balance, $V_{\rm s,cloud} \approx V_{\rm
s,IGM}\,\sqrt{n_{\rm IGM}/n_{\rm cloud}}$ (Mendoza 2000), and so for a
typical mean shock velocity of the cocoon of 0.02$c$\footnote{Note that
this mean velocity of the shock front is lower than the velocity at
which the radio size increases, because the transverse radio shocks are
slower by a factor of the aspect ratio of the radio source, typically
3--5. However, due to the $v^3$ dependence the actual value assumed is
decreased by a smaller factor than this.}, the velocity of the shock
within a cloud is $\sim 200\,{\rm km\,s}^{-1}$. This is within the
range of values considered by the Dopita and Sutherland models.

The shocked gas cools on a time-scale of about $10^4$ years according
to the Dopita and Sutherland models, very much shorter than the
lifetime of a radio source. The direct ionization of the cool clouds
by the passage of the these shocks will therefore be a short--lived
effect, and the associated boosting of emission line luminosities will
only be relevant for those clouds close to the bow shock.  For a radio
source of about 50\,kpc size, assuming emission line clouds of density
$n_{\rm cloud} \sim 10^8\,\rm{m}^{-3}$ with a volume filling factor of
$\sim 10^{-6}$, the total radiative flux produced in the clouds by the
direct passage of the shocks will be of order $10^{34}$\,W.  When
compared to the [OII] emission line luminosities of $10^{35}$ to
$10^{36}$\,W (which constitutes the majority of the cooling) it is
clear that these direct shocks are not the dominant  
source of ionization.  Further proof of this is that the emission line
regions do not trace out the shock front, but seem to extend throughout
the volume of the cocoon.

Although direct ionization of the cool clouds by shocks cannot
adequately explain the observations, the hot phase has a filling
factor of unity, and the properties of the emission line regions may
be better explained by the UV photon field produced by the cooling
post--shock hot gas at the edges of an expanding radio
source. Assuming that the hot gas obeys the scaling relations of
Dopita \& Sutherland, entire shells of the hot phase material can
become radiative and act as a photon source for the cool phase. 
These photons travel both upstream and downstream
from the shock, creating a precursor H\textsc{ii} region upstream and
strongly influencing the temperature and ionization state of the
post-shock gas. Indeed, the location of the 6C galaxy data on the line
ratio diagnostic diagram (Fig.~\ref{Fig: 14}) indicates that those shock
models which include a precursor ionization region provide a better
description of the ionization state for the small radio sources.

The UV luminosity of the hydrogen ionizing radiation field is given by Dopita and
Sutherland (1996) as
\begin{equation*}
L_{\rm UV}  = 1.11 \times 10^{-6}\,(V_{\rm s}/100\,{\rm km}\,{\rm s}^{-1})^{3.04}\,(n/10^6{\rm m}^{-3})\,{\rm W}\,{\rm m}^{-2}.
\end{equation*} 
The expanding radio cocoon can be approximated as an expanding sphere
centred on the AGN.  Before the shock front reaches a given cool,
dense gas cloud, the flux of ionizing photons received from the shock
as it passes through the hot, diffuse gas will be roughly the same as
that of a point source with the same total luminosity as the shock.
Beyond this point, the evolution (with increasing size of the radio
source) of the flux of ionizing photons from the shock which reach the
cloud then depends upon how the density of the external hot, diffuse
gas decreases with distance from the host galaxy, and how the velocity
of the bow shock varies with radio source size.

In a study of FRII galaxy environments, Wellman, Daly \& Wan (1997)
found that the variation of the density of the hot diffuse gas
surrounding radio galaxies could be well fitted by King profiles of
the form   
\begin{equation*}
n(r) = n_0[1+(r/r_c)^2]^{-3\beta/2},
\end{equation*}
with central density $n_0 \sim 3\times10^4\,{\rm m}^{-3}$, core radius
$r_{\rm c} \sim 50-75 {\rm kpc}$ and $\beta \sim 0.7$.  Studies of
radio galaxies and clusters by Neumann (1999), Siebert, Kawai \&
Brinkmann (1999) and Heinz, Choi, Reynolds \& Begelman (2002) have
found similar values for these parameters. The density of the emission
line gas can therefore be approximated as being roughly constant
density within the core radius (50\,kpc) and decreasing proportional
to $r^{-2}$ significantly beyond this.  Kaiser \& Alexander (1997)
predict that, in the self-similar expansion of the cocoons, if the
environmental density can be approximated as a power--law, $n \propto
r^{-\gamma}$, then the radio source grows with time as: $D_{\rm rad}
\propto t^{3/(5-\gamma)}$.  For small radio sources ($\gamma = 0$),
the expansion velocity of the radio source bow shock therefore depends
on the source age and/or size as: $v \propto t^{-2/5} \propto
D_{\rm{rad}}^{-2/3}$. At the other extreme, the expansion velocity of
the largest radio sources in the sample will be approximately constant 
(since $\gamma \sim 2$).  We have combined this variation in expansion
velocity with radio source size with a King profile density model.
Integrating over the total surface area of the shock front, the total
output energy from the expanding shock front remains constant over
time as it passes through the diffuse gas.  The flux of UV photons
received by a cool, dense gas cloud from this expanding shock front is
roughly constant prior to the shock front reaching the cloud.  Once
the shock has passed through a cloud, the flux of photons received by
the cloud decreases as $D_{\rm{rad}}^{-2}$ as the radio source grows. 

Using the scaling relations of Dopita \& Sutherland and a mean shock
velocity at 12\,kpc of 0.02$c$ in the diffuse IGM surrounding the
clouds, the precursor luminosity per unit area of the shock front
varies with radio source size from over $10^{-3}\,{\rm W}\,{\rm
m}^{-2}$ within 20\,kpc from the AGN to $\sim 10^{-4}\,{\rm W}\,{\rm
m}^{-2}$ at a distance of roughly 100\,kpc. Our observations indicate
that the extent to which we can observe a purely photoionized emission
line region is about 25 kpc from the host galaxy.  A cloud at this
distance would receive a flux density of $\approx 6.7 \times
10^{-4}\,{\rm W}\,{\rm m}^{-2}$ from a typical point source of
luminosity $5 \times 10^{39}\,{\rm W}$ (McCarthy 1993).  This is
approximately the same as the flux density of a shock at about 40 kpc.
For different values of the shock velocity, ambient gas density and
AGN power, shocks can provide a greater or lesser flux of ionizing
photons than the central engine.  It is quite feasible that the
precursor field produced by the shocks associated with the expansion
of the radio source can ionize the emission line regions out to
distances of 50-60\,kpc. 

\begin{figure}
\vspace{2.45 in}
\begin{center}
\includegraphics{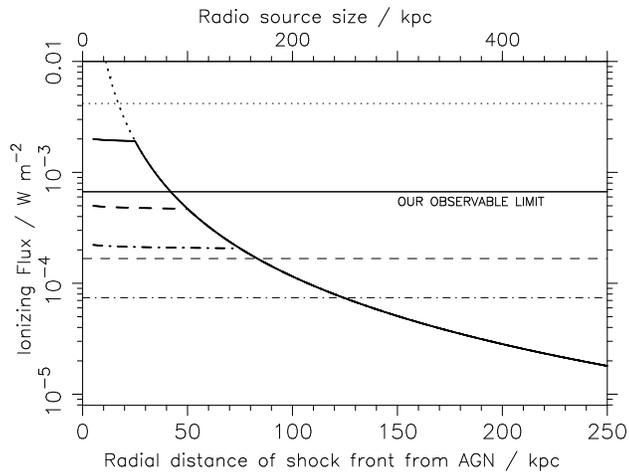}
\end{center}
\caption{The changing contribution of ionizing photons from shocks and
the AGN with radio size.  Four clouds are considered, located at
distances of 10kpc, 25kpc, 50kpc and 75kpc from the AGN.   Horizontal
lines represent the constant flux received from the $5 \times
10^{39}\rm{W}$ AGN by each cloud.  The contribution from the ionizing
shock varies with time, and the changing flux received by each cloud
is denoted by the four remaining lines.  Dotted tracks model the flux
received by the cloud at 10kpc, solid tracks model the flux received
by the cloud at 25kpc, dashed tracks model the flux received by the
cloud at 50 kpc and dot-dashed tracks model the flux received by the
cloud at 75 kpc.  The horizontal solid line also represents the
limiting flux of photons required by the clouds at a given radius in
order for them to be observed at the sensitivity of our
observations.  For large radio sources, the observed maximum extent of
apparently  purely photoionized regions was $\sim 25$kpc; this line
therefore coincides with the line representing the flux 
received by a cloud at 25kpc from a typical AGN. 
\label{Fig: 21}}
\end{figure}

This model is illustrated in Fig.~\ref{Fig: 21}, which shows the
changing contribution of ionizing photons to gas clouds at different 
distances from the AGN as the radio source expands.  For all but the
very smallest radio sizes, the ionization state of a cloud at 10kpc is
dominated by photoionization by the AGN.  However, for radio sizes $<
10$kpc our assumed values for $V_{\rm s}$ and $n$ are not valid, as
the presence of the host galaxy is not included in our model.  The
ionization state of a cloud at 25 kpc is dominated by the effect of
shocks when the radio source is small (20\,kpc $\lta D_{\rm{rad}} \lta
$70\,kpc).  Emission from a cloud at 50 kpc from the AGN would be
shock dominated out to radio sizes of $\sim 150$kpc, but would only be
observed in our data out to radio source sizes less than or comparable
to the size of the EELR; the combined flux of ionizing photons from
the shock and the AGN is approximately the same as that from the AGN
at a distance of 25\,kpc.  The emission from such a cloud will be
dominated by shock ionization.  Clouds at a greater distance from the
AGN than this will not be observable in this model at the sensitivity
of our observations.   

We note that the expanding shock front is not strictly spherical, as
considered in this model, but propagates at different velocities at
different locations on the cocoon.  The shock will be most luminous
near the radio source hotspots, by a factor of $\sim$ 30-60 for radio
sources with $D_{\rm{rad}} <120\,{\rm kpc}$. This will enhance the
alignment effect.  An alternative source of ionizing shocks are
interactions between individual gas clouds and the radio jet. Although
this would lead to shocked gas clouds closely aligned with the radio
axis, the variation in EELR size with radio source size is much more
difficult to explain. 

An important question is whether or not such luminous shocks are
permitted by UV and X-ray observations of radio galaxies.  The total
number of ionizing photons emitted from the radio sources bow shocks
is comparable to that of a luminous quasar; however, the spectral
energy distribution of the shock waves is considerably steeper in the
UV.  The shock models of Dopita \& Sutherland (1996) predict a flux
density of $4 \times 10^{-21}\,{\rm W}\,{\rm m}^{-2}\,\rm{\AA}^{-1}$
at a rest-frame wavelength of $1800$\AA\, for the shock parameters
considered above.  The flux densities of 3CR sources at $z \sim 1$ are
$\sim 5 \times 10^{-21}\, {\rm W}\,{\rm m}^{-2}\,\rm{\AA}^{-1}$ in the
rest-frame waveband 1670-1890\AA\, as observed through a 1.5 arcsec
slit (Best {\it et al} 2000a).  Since these observations include only
a fraction of the shocked gas, the shock models cannot be excluded.
Observations of higher redshift radio sources are also in agreement
with this result; the predicted flux density of the shock models do
not exceed the continuum flux density observed in the spectra of 6C
radio sources at redshifts out to $z > 2$ (Jarvis {\it et al} 2001).

The soft X-ray emission from the shock falls off rapidly with photon
energy.  For the parameters of our shock model, the shock front would
have a flux density $\sim 10^{-7}\,{\rm W}\,\rm{m}^{-2}$ for $h\nu >
0.5\,\rm{keV}$.  X-ray observations of 3C radio sources (e.g. Allen
{\it et al.} 2001; Worrall {\it et al.} 2001) have found upper limits
to the X-ray flux density of about 3-10 times greater than this value.  The
observations often show holes in the X-ray emission, coincident with
the lobes of the radio source.  These cavities are surrounded by
bright X-ray emission, which can be modelled as limb brightened shells
of gas formed by the displacement and compression of material by the
expansion of the radio source. The brightest X-ray emission from 3C
317 (Blanton {\it et al} 2001) is coincident with the H$\alpha\, +$
[N\textsc{ii}] contours of Baum {\it et al.} (1998), providing
evidence for cooler gas in these regions as well. Note that it is the
hot inter-cloud gas which emits at X-rays, rather than the cooler gas
clouds. If the cool clouds were heated to temperatures $T \ge 10^6$ K,
the gas would not be able to cool by line emission and no line
emission would be observed, regardless of whether the gas was ionized
by shocks or the AGN. 

In general, then, this model fits the observational data very well.
For the very smallest radio sources, it is expected that the flux of
ionizing photons from the AGN is of far greater importance than those
produced by any ionizing shocks associated with the expanding radio
cocoon.  Jarvis {\it et al.} (2002) observe no shock boosting of the
Lyman alpha line in small radio sources, and Hirst, Jackson and
Rawlings (2002), provide tentative evidence that \ooo\, is not boosted
in Compact Steep Spectrum (CSS) sources, consistent with this
model.  The observed reduction in 
Balmer line flux of the small radio sources confirms the presence of
ionizing shocks in these sources (Morse {\it et al.} 1996), rather
than indicating a reduction in ionizing photons.  Further, Moy \&
Rocca-Volmerange (2002) have considered the predictions of combined
shock ionization and photoionization models.  Comparing the emission
line ratios of a large number of radio galaxies and quasars, they also
find that both ionization processes are likely to be occurring in the
emission line regions of galaxies with radio sizes 2\,kpc $<
D_{\rm{rad}} < 150$ kpc, with pure photoionization by the AGN
remaining the best interpretation for the emission line regions of the
smallest and largest sources.

An interesting aspect of this model is that if a cloud at 50\,kpc radius
can be ionized by shocks as they approach this radius, that cloud should,
in principle, also be ionized by shocks when the radio source has much
smaller radii. In practice, however, the line emission is generally not
seen much beyond the limit of the radio emission. At first sight this
seems to argue against such strong shock contributions. However, even in
the smallest sources there is no observable emission from much beyond the
radio source, in regions where photoionization should be occurring. This
suggests that the lack of line emission from these regions is not
necessarily a problem with the shock models, but rather a general feature
of radio sources that the emission line regions become much more luminous
after the passage of the radio sources. Further evidence in support of
this comes from observations that the emission line regions around radio
quiet quasars are not significantly extended (Telfer et al 2000) whilst
those of radio loud quasars show extensions comparable to those of radio
galaxies (Crawford and Vanderriest 2000).

Clearly, regardless of the ionizing photons that it supplies, the
expanding radio source somehow directly increases the emission line
observability of regions that it has passed through. One possibility is
that the gas clouds in the EELR are dragged out from a more central region
of the galaxy by the radio source for a limited distance and/or time, and
so are only present at these large radii after the passage of the
radio source. However, our modelling shows that clouds at distances of
$> 25\rm{kpc}$ from the AGN \emph{can} be in place prior to the expansion of
the radio source and yet not be observed, and that clouds at a distance
greater than 60kpc from the AGN cannot be ruled out.  The presence of
extended Lyman-$\alpha$ halos around radio galaxies (e.g. van Ojik {\it et
al} 1997) provides further evidence that dense gas clouds may exist out to
large radii around radio sources of all sizes. An alternative possibility
is that the passage of the radio jet disrupts large, optically thick
emission line clouds, producing smaller optically thin clouds which
increases the covering factor for absorption of AGN radiation (cf. Bremer
et al 1997). This appealing possibility also helps to explain the radio
size evolution of the optical alignment effect.

\subsection{Emission line region kinematics}

The velocity FWHM of the emission line gas is strongly anticorrelated
with radio size for both 6C and 3CR galaxies. This is interpreted as
the boosting of the observed FWHM of small sources by shock
acceleration and entrainment of the emission line gas clouds over that
observed in the larger photoionized sources.  The correlation of both
the kinematic and ionization properties of the radio galaxies with
radio size within both samples indicates that these are related,
regardless of the difference in radio power between the 3CR and 6C
galaxies.  For the 6C sources, the range of velocities observed
displays a weaker correlation with emission line region size and FWHM
than the more powerful 3CR sources.    

One difference between the two samples is the range of velocities
observed in the sources, which appears to decrease with both radio
power and radio size.  The weaker 6C radio sources display a somewhat
lesser degree of variation in the emission line gas kinematics than is
seen among the 3CR galaxies.  This is not likely to be caused by any
effects of the lower S/N of the 6C observations; no evidence is seen
in Figs.\,1-10 that the fitting procedures are missing further
velocity components at greater spatial extents than the regions 
currently analysed.    If confirmed by a larger data set, the
variation in velocity range with radio power could have several 
different explanations.    

Large radio sources at $z \sim 1$ have velocity profiles consistent
with rotation (Best {\it et al} 2000b), as do 3CR sources at low
redshifts (Baum {\it et al.} 1992).   $K$-band observations show that
$z \sim 1$ 6C radio galaxies are fainter by $\sim 0.6$ magnitudes
(Eales {\it et al} 1997, Inskip {\it et al} 2002) than 3CR galaxies at
the same redshift.  If this is due to a difference in host galaxy mass, the
rotation profiles of the less massive 6C galaxies would be a factor of
$\sim 1.3$ less.  The velocity profiles could alternatively indicate
infall or outflow of material.  For small radio sources, the extreme
gas kinematics observed may be due to shocks associated with the radio
jet and/or the expansion of the radio source, and would therefore be
strongly correlated with radio jet kinetic power, and hence the
luminosity of the radio source.  The mean velocity range of small 3CR
galaxies is 45\% greater than that observed for small 6C galaxies,
although this result is based on a very small number of sources. The
more extreme kinematics of the 3CR sources are as expected, given
their greater radio power.   

\section{Conclusions}

Very deep spectroscopic observations have been made of an unbiased
subsample of 8 6C galaxies at $z \sim 1$. Many emission lines have
been observed over the rest-frame wavelength range 1500-4500\AA, and a
study of the two-dimensional kinematics of the emission line gas has
been carried out.  Our conclusions can be summarised as follows:    

\begin{enumerate} 
\item The observed spectra of the 6C galaxies are quite varied, both
in the strength of the emission lines observed, and the line ratios
observed.  
\item The composite spectra of the 6C galaxies are similar to those of
Best {\it et al} (2000a) for the 3CR sources at the same redshift.   
\item 6C sources with $D_{\rm{rad}} > 120\,\rm{kpc}$ host less
powerful AGN than 3CR sources of a similar size at the same redshift.
Their spectra are well explained by photoionization, typically with a lower
ionization parameter than their more powerful 3CR counterparts.  The
total emission line luminosities of large 6C sources are also smaller
than that of large 3CR sources, and their emission line regions are
observable out to smaller physical scales.   
\item For small radio sources, $D_{\rm{rad}} < 120$\,kpc, a
combination of AGN photoionization and shock ionization provides the
best explanation of their spectra.  Their emission line regions
typically have a similar size to the extent of the radio source.  
\item The velocity profiles of shock ionized EELRs are distorted,
whereas photoionized EELRs display smooth velocity profiles.  
\item The ionization properties of the subsample can be explained by a
simple model incorporating photoionization by the AGN and a luminous
shock associated with the expanding radio source. 
\item The kinematics of the EELRs of 6C radio sources are similar to
those of more powerful 3CR sources. 
\item A high velocity component is observed in the EELR of 6C1019+39
at $\sim 700 \rm{km\,s}^{-1}$, close to the host galaxy.  
\end{enumerate}

In summary, the properties of 6C radio galaxies at $z \sim 1$ are
similar to 3CR sources at the same redshift, despite the decrease in
radio power between the samples.   A strong anticorrelation of
emission line luminosity with source size is found.  In addition, we
also find tentative evidence that the range of velocities observed in the
emission line gas is most likely dependent on the properties of the
radio source rather than the underlying gravitational potential of the
host galaxy.   In a companion paper (Paper 2), this dataset is
compared with lower redshift sources of the same radio power to break
the radio power-redshift degeneracy, and to investigate the intrinsic
dependencies of radio galaxy properties on redshift, radio power and
source age.  

\section*{Acknowledgements}

This work was supported in part by the Formation and Evolution of
Galaxies network set up by the European Commission under contract ERB
FMRX--CT96--086 of its TMR programme. 
KJI acknowledges the support of a PPARC research studentship.  
The William Herschel Telescope is operated on the island of La Palma
by the Isaac Newton Group in the Spanish Observatorio del Roque de
los Muchachos of the Instituto de Astrof{\'{\i}}sica de Canarias.  
This research has made use of the NASA/IPAC Extragalactic Database
(NED) which is operated by the Jet Propulsion Laboratory, California
Institute of Technology, under contract with the National Aeronautics
and Space Administration.  PNB is grateful for the generous support
offered by a Royal Society Research Fellowship.
We would like to thank the referee for useful comments.

\appendix
\section{Faint broad components in the [O\textsc{ii}] 3727\AA\, line of 6C and 3CR galaxies}
In light of the results of Sol\'{o}rzano-I\~{n}arrea {\it et al}
(2001), as described in the introduction to this paper, we have
undertaken a search for such underlying broad components in the \oo
3727\AA\, lines of the 6C subsample, and also in the spectroscopic data for
the 3CR subsample of Best {\it et al} (2000a, 2000b).  In the main body of
the paper, previous fitting of a two-dimensional region around the
\oo\, emission line used a series of extracted spectra four pixels in
width, stepped every two pixels, with the purpose of investigating the
spatial variations of the emission line properties.   As already noted
in section 3.1, a single Gaussian fit to the emission line data was
inadequate in several cases. The peak flux was often underestimated
for these sources, coupled with a poor fit at high velocities, as
expected for mis-fitting the combination of a narrow line with a faint
broad component.  An alternative explanation is that the line profiles
may be non-Gaussian.

We have used a similar process for fitting weak broad components to
the emission line using a single wider spectrum extracted from the
centre of each source, thus increasing the signal--to--noise over that
of the previous extracted spectra.  The new extracted region was
typically 9 pixels (3.24 arcsec) in spatial extent.  A wider extracted
region could have been used, but in such a situation any velocity
gradient present in the 2--dimensional spectrum would mimic an
underlying broad component in the data.  The same constraints were
used as for narrow line fitting, with the additional restrictions that
any broad component should be fainter and have a wider FWHM than the
first Gaussian fitted to the data.  

Of the 22 6C and 3CR sources studied, only six showed clear evidence
for an underlying broad component, with a single Gaussian (or two
Gaussians for the sources with more than one distinct component)
providing a larger reduced $\chi^2$ than that achieved with the
addition of another Gaussian component.  These were 3C22, 3C217,
3C247, 3C265, 3C280 and 3C441.  Several other sources were also poorly
explained by a single Gaussian fit, however in these cases additional
components gave no improvement in the reduced $\chi^2$.  The
two-Gaussian reduced $\chi^2$ for 6C1017+37 was comparable to that for
a single Gaussian; for this source the presence of an additional
component is uncertain.  The fact that 6 3CR sources showed evidence
for broad components whilst none of the 6C sources did is not
necessarily significant since the presence of an underlying broad
component was rejected for sources for which the 2-Gaussian reduced
$\chi^2$ was less than that for the single Gaussian fit and the lower
signal--to--noise of the 6C data is an important detriment to
improving the fit.   

We found that in the instances for which a broad component is
required, several different combinations of two or more Gaussians had
comparable reduced $\chi^2$. As well as the expected bright narrow
line with a weak underlying broad component, the data was also
generally well fitted by a somewhat fainter narrow Gaussian coupled
with a strong, broad component. The fairly low S/N of the data allowed
many possible groups of Gaussians to fit the data well within
experimental error, and so we are unable to determine exactly the
parameters of any actual underlying broad component in the emission
lines of the sources studied.  We can, however, gain a greater
understanding of the inadequacies of single-Gaussian fitting, which
generally underestimates the peak of the line, and misfits the shape
of the wings.  We combined the emission line data of the six sources
showing evidence for some underlying broad component, by centering
their peaks and re-binning the data to the same FWHM. A single
Gaussian was fitted to the resulting combination; this and the
residuals obtained are plotted in Fig.~A1. This clearly shows how a
single Gaussian misfits the data. 

\begin{figure}
\vspace{4.5 in}
\begin{center}
\includegraphics{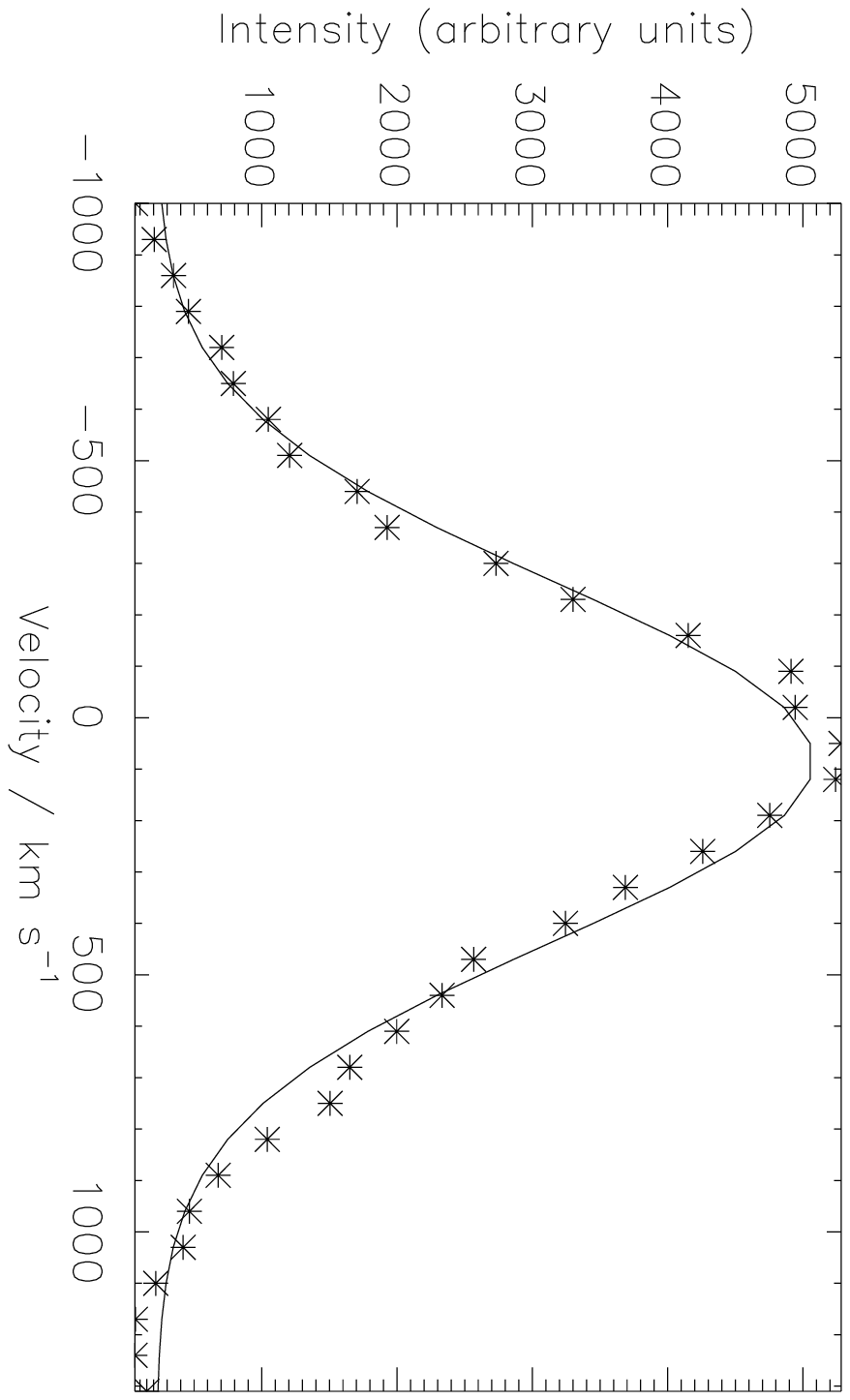}
\includegraphics{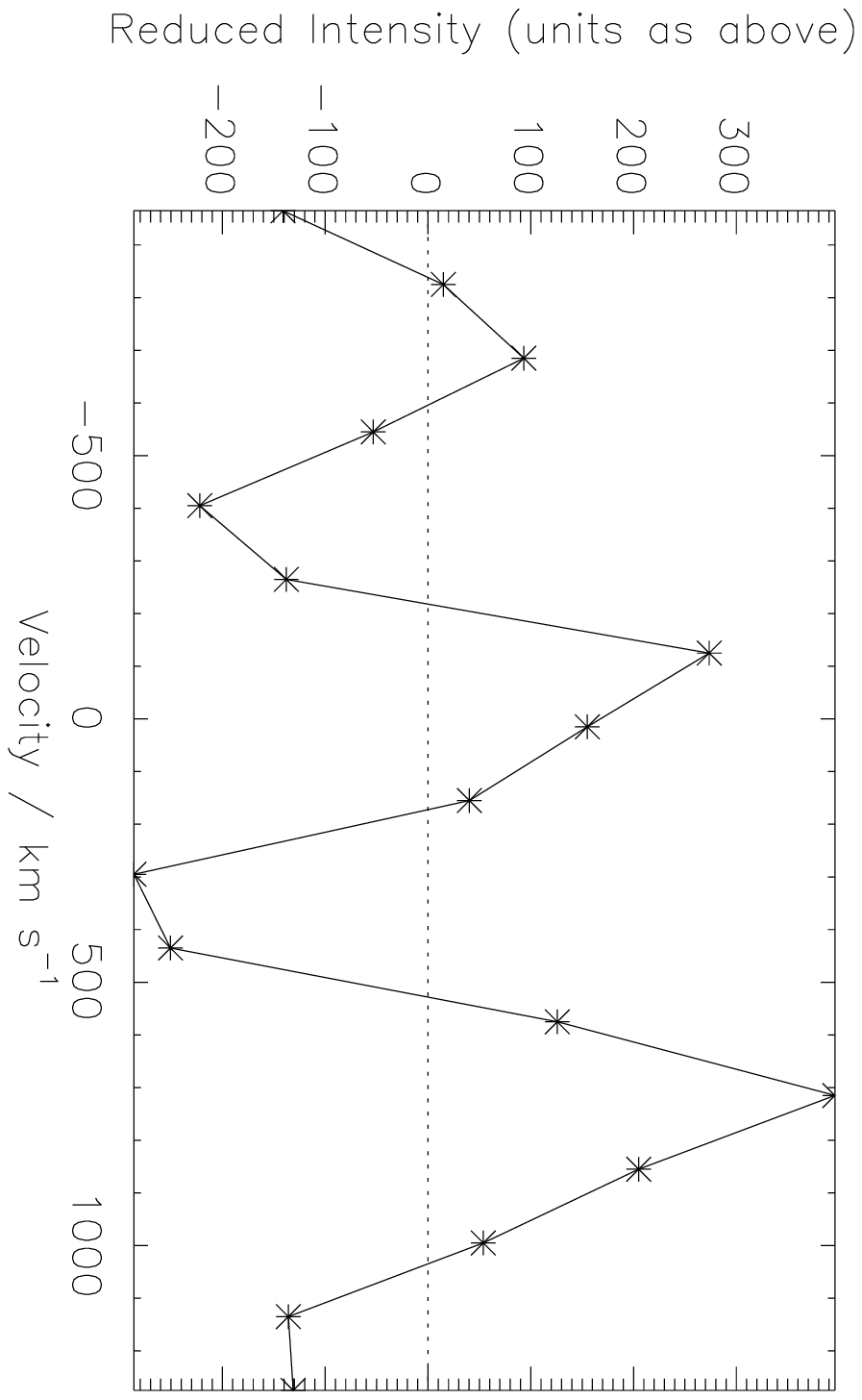}
\end{center}
\caption{Searching for underlying broad components in the \oo 3727\AA\,
emission line. (a) Single Gaussian fit to the extracted line flux for
six sources showing evidence for an underlying broad component. (b)
Residuals of the 1--Gaussian fit.
\label{Fig: A1}}
\end{figure}

On the whole, the necessity for broad components is seen to be more
likely in intermediate--to--small radio sources. This is certainly in agreement with
the shock--related origins of broad components, as larger sources are
in general seen to be photoionized.  On the other hand, the very smallest
sources do not exhibit clear evidence for broad components in their
emission lines. However, the complex shock--induced velocity
structures of these sources, already fitted by multiple Gaussians,
will inhibit the fitting of further velocity components. Thus, whilst
broad components may indeed be present in these heavily shocked
smaller sources, they cannot be conclusively identified from these data.
The work of Sol\'{o}rzano-I\~{n}arrea {\it et al} benefited from a
higher S/N than our data, as their observations were made with a much
lower spectral resolution.

\label{lastpage}

\end{document}